# Chirality-induced magnetoresistance in hybrid organic-inorganic perovskite semiconductors

Md Azimul Haque,[1,2] Pius Markus Theiler,[1] Ian A. Leahy,[1] Steven P. Harvey,[1] Jeiwan Tan,[1] Matthew P Hautzinger,[1] Margherita Taddei,[1] Aeron McConnell,[3] Andrew Greider,[4] Andrew H. Comstock,[3] Yifan Dong,[1] Kirstin Alberi,[1] Yuan Ping,[4,5,6] Peter C. Sercel,[7] Joseph M. Luther,[1,2] Dali Sun,[3] Matthew C. Beard[1,2]*

[1]National Laboratory of the Rockies, Golden, CO, USA
[2]Renewable and Sustainable Energy Institute, University of Colorado Boulder, Boulder, CO, USA
[3]Department of Physics and Organic and Carbon Electronics Lab (ORaCEL), North Carolina State University, Raleigh, NC, USA
[4]Department of Materials Science and Engineering, University of Wisconsin-Madison, Madison, WI, USA
[5]Department of Physics, University of Wisconsin-Madison, Madison, WI, USA
[6]Deparment of Chemistry, University of Wisconsin-Madison, Madison, WI, USA
[7]Center for Hybrid Organic Inorganic Semiconductors for Energy, Golden, CO, USA
*Corresponding author. Email: matt.beard@nrel.gov

**Abstract**

The combination of semiconducting properties and synthetically tunable chirality in chiral metal halide semiconductors (CMHS) offer a compelling platform for room temperature control over electronic spin properties, leveraging effects such as chirality-induced spin selectivity (CISS) for the development of new opto-spintronic functionalities. We report room-temperature CISS-induced magnetoresistance (CISS-MR) exceeding 100% for spin valves in a configuration consisting of a ferromagnet (FM), tunneling barrier, and CMHS. The high CISS-MR is attributed to interfacial spin-selective tunneling barrier induced by the chirality, which can produce current dissymmetry factors ($g_c$) that surpass the limit imposed by the Jullière model governed by the intrinsic spin polarization of the adjacent FM contact. The CISS-MR exhibits a strong dependence on the CMHS composition, revealing a structure-property relationship between CISS and structural chirality. The observed exceptionally large tunneling MR response differentiates from a subtle anisotropic MR arising from the proximity effect at the FM/CMHS interface in the absence of a tunneling barrier. Our study provides insights into charge-to-spin interconversion in chiral semiconductors, offering materials design principles to control and enhance CISS response and utilize it in functional platforms.

Metal halide semiconductors (MHS) are widely studied for optoelectronic applications due to their tunable semiconducting properties. The hybrid organic/inorganic nature of MHS enables properties that are not available to either purely organic or inorganic systems. The prototypical realization of such emergent properties is in the incorporation of chiral organic cations, producing a family of enantiomerically pure chiral semiconductors (CMHS) with novel functionality whose properties can be tuned through harnessing the synergistic interactions between the chiral organic and inorganic components. Enabling phenomena such as the circularly polarized galvanic effect, large spin hall effect, the chiral-phonon-activated Seebeck effect, and the chirality-induced spin selectivity (CISS) effect.(*1-5*)

CISS enables control over spin populations at room temperature without requiring magnetic fields or ferromagnets providing a fundamentally new route to achieve spin functionality.(*6*) The CISS effect, when realized in a chiral semiconductor, forms the basis for various chiral-optoelectrical applications such as circularly polarized light detection, spin-light emitting diodes, and chiral spin valves.(*7-9*) A spin valve is a simple two-terminal device that exhibits two distinct resistance states depending on the relative alignment of its magnetic layers, termed magnetoresistance (MR). Unlike conventional spin valves requiring two ferromagnetic (FM) electrodes, or a FM and a hard magnetic electrode, CISS-based spin valves operate with a single soft FM and the two resistance states are determined by the alignment between the magnetic polarization and stereochemical configuration. As a result, CISS-based spin valves operate through a new mechanism (chirality) whose understanding and control could pave the way for new unrealized functionality not available using traditional approaches. While CISS has been observed in DNA, peptides, self-assembled monolayers, and chiral quantum dots,(*10-12*) understanding its structure-property relationships remains challenging due to the limited tunability of existing CISS-active thin films and general fabrication methods. CMHS with composition $A_2BX_4$, where A is an organic chiral cation, B is the metal cation, X is a halide anion provide an ideal platform to overcome this limitation, offering synthetic control over semiconducting properties via B and X sites, and symmetry broken crystal structures via A-site chirality, making them promising candidates for CISS-based applications and developing an understanding of the underlying CISS mechanisms.(*13-16*).

Here we developed a chiral spin valve based upon a spin-dependent tunnelling junction which exhibit CISS-induced MR (CISS-MR) well in excess of 100% at room temperature, comparable to conventional spin-valves, which exceed ~200% in consumer electronics.(*17*). We elucidate how the spin-dependent tunnelling barrier, induced by CISS, can achieve MR values consistent with tunneling magnetoresistance (TMR) and show that without the tunnelling barrier CISS-MR values are much smaller. The interfaces between the tunnelling barrier, FM, and chiral semiconductors play the key role in achieving high MR values reported here.  With this platform we established structure/function relationships by comparing the MR from four different chiral A-site cations that induce varying degrees of chirality, assessed through the continuous chirality measurement (CCM) and employed into three different spin valve architectures. Our findings provide design principles to enhance and understand charge-to-spin interconversion in chiral semiconductors, paving the way for disorder-tolerant, high-MR spin functionality. Similar to the success envisioned with solution-processed halide perovskites in potentially revolutionizing solar cell technology, CMHSs offer a unique opportunity to merge the structural tunability of perovskite semiconductors with the intrinsic spin selective functionality inherent in chiral systems.(*18*)

**Spin valves fabrication and analysis**

The CISS effect describes a process whereby a charge current becomes spin polarized along the current direction when passing through a chiral medium (**Fig. 1a**). A traditional spin valve typically consists of a hard magnet and a FM that sandwiches a normal metal or insulating tunnel barrier. The efficacy of the spin injection and transport in such structures is assessed based on the magnetoresistance (MR) which is defined as

$$MR\% = \frac{G_P - G_{AP}}{G_{AP}} \times 100\%, \qquad [1]$$

where $G_P$ is the conductance when the FM is aligned with the hard magnet, and $G_{AP}$ is the conductance when they are antiparallel. In a CISS-based spin valve, one of the magnetic electrodes, typically the hard magnet, and the non-magnetic interlayer are replaced by the chiral semiconductor (**Fig. 1b**). The CISS layer assumes the same role as the hard magnet (i.e. it is fixed with respect to soft ferromagnetic and applied magnetic field). The CISS-MR can therefore be similarly defined where now $G_P$ is the conductance when the chirality and FM are parallel (**Fig. 1c**), i.e. the device is in the low resistance/high conductance state, while $G_{AP}$ is when it is in the high resistance/low conductance state (**Fig. 1d**). Here we adopt the notation and definitions described by Weiss et. al.(*19*) Commonly, in CISS measurements the current dissymmetry factor $g_c$ rather than MR is assessed using a magnetic conducting atomic force microscopy (mc-AFM) measurement where a nanoscale current, $I_{M\downarrow,\uparrow}$, is measured under opposite magnetization orientation of the FM, using

$$g_c\% = \frac{I_{M\uparrow} - I_{M\downarrow}}{I_{M\uparrow} + I_{M\downarrow}} \times 100\%. \qquad [2]$$

This quantity represents the spin-dependent charge current polarization measured in a transport experiment.(*19*) Researchers have reported a broad range of $g_c$ values using the mc-AFM technique for CMHS films, and they are typically in the range of 50-94%, indicating an efficient spin-polarization that should also correspond to a large MR, since the MR and $g_c$ are related by $MR = 2|g_c|/(1 - |g_c|)$. However, despite the strong spin polarization inferred from these mc-AFM measurements achieving apparent MR values in the hundreds of percent, MR values in reported macroscopic CISS-based spin valves remain low.(*3, 9, 20-22*) This discrepancy highlights unresolved design rules for translating the high spin-polarization observed in typical CISS demonstrations to the practical implementations of CISS spin valves.

Previous reports of macroscopic CMHS spin valves have utilized buffer layers between the CMHS layer and non-magnetic top metal electrode to avoid unintended surface reaction and potential short circuits.(*23*) There are also reports where no such buffer layers are employed.(*9*) The most demonstrated CISS-based spin valves do not have such buffer layers.(*24*) This calls into question what is the best architecture for CMHS-based spin valves to pursue high MR response in practical CISS applications. To resolve this important question, we investigated three distinct spin-valve architectures by varying the number of buffer layers, i.e. spin-valves with two, one, and no buffer layers between the CMHS and non-magnetic electrode. Supplementary Fig. 1 shows schematics of these architectures and that the highest MR value is observed for spin valves with no buffer layers. This result indicates that reducing the number of interfaces suppresses spin trapping and scattering. A typical spin valve architecture (FM/Al$_2$O$_3$/CMHS/Au) is shown in **Fig. 1b,** here an ultrathin (2 nm) Al$_2$O$_3$ tunneling barrier is employed between the FM and CMHS layer. Au is used as the top electrode since its work function is closer to the valence band of the CMHS. Upon the application of an out-of-plane magnetic field, the spin valve exhibits either a low or high resistance state (**Fig. 1c-d**) depending on the CMHS chirality and applied field direction. **Fig. 1e-f** presents the SEM images of the top view of the macroscopic spin valve and the schematic cross-sectional view of the full stack. We discuss later what happens in the absence of the tunneling barrier where a smaller MR, but larger current can be observed.

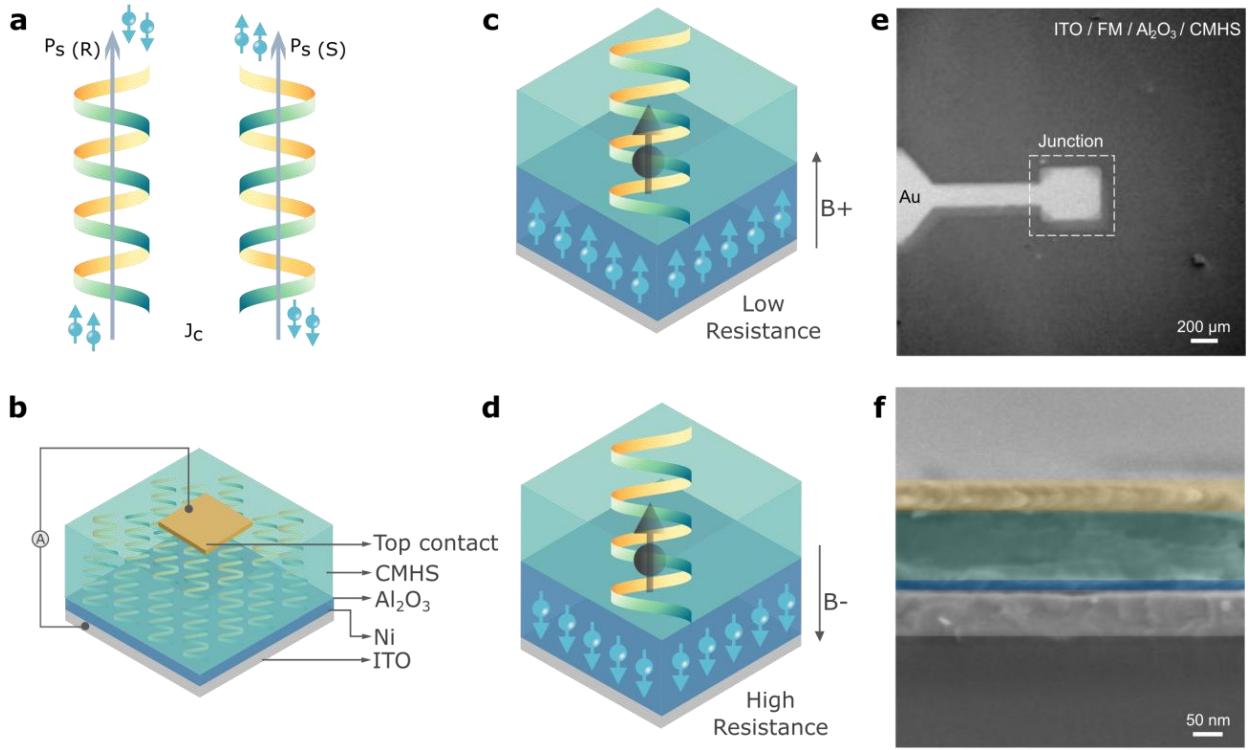

**Fig. 1| CISS-based spin valve design and response. a**, Schematic illustration of the CISS effect, where a flow of charge current $J_c$ in the enantiomorphs generates a pair of spin polarizations $P_S$. **b**, CISS spin valve architecture. CMHS enantiomorph with a preference for spin-up carrier transport separated from Ni electrode by Al$_2$O$_3$ tunnel barrier. A low resistance **c**, state is observed when the Ni and enantiomorph spins are parallel and high resistance **d**, when the spins are antiparallel. SEM images showing **e**, top view of the spin valve and **f**, cross section of the spin valve with the architecture ITO/Ni/Al$_2$O$_3$/CMHS/Au.

With this structure (FM/Al$_2$O$_3$/CMHS/Au) held constant, we first studied spin valves using the canonical (*R/S*-MBA)$_2$PbI$_4$. Typical room-temperature current-voltage (*I-V*) characteristics (**Fig. 2**) of the spin valve show a non-linear response, indicating the formation of a tunnel junction.(*25*) For the (*R*-MBA)$_2$PbI$_4$ spin valve (**Fig. 2a**), a lower resistance/higher conductance is observed for the positive field, +B, compared to that for the negative field, -B, exhibiting a preferential transport of carriers with spin orientation parallel to the positive magnetic field. The *I-V* characteristics exhibit an opposite behavior for the (*S*-MBA)$_2$PbI$_4$ spin valve (**Fig. 2b**). There is a lower resistance/higher conductance for -B compared to +B (also see Supplementary Fig. 2). Such opposite but symmetric spin-dependent *I-V* characteristics from (*R*-MBA)$_2$PbI$_4$ to (*S*-MBA)$_2$PbI$_4$ coincides with that expected from CISS-MR phenomena.(*26*) Reversing either the structural chirality or the external magnetic field orientation switches the conductance state from high to low or vice versa, while reversing both the chirality and magnetic field yields similar *I-V* characteristics. A histogram of 106 tested spin valves for various CMHS compositions (**Fig. 2c,** Supplementary Fig. 3) show that most of the fabricated spin valves in this study exhibited an MR response in the 20-50% range (**Fig. 2a** is a representative 50% MR and **Fig. 2b** exhibits MR of 20%) with an average response already showing a large improvement over literature results and more in-line with the mc-AFM results. Detailed fabrication guidelines for achieving reliable spin valves are discussed in Supplementary Information (Supplementary Note 1 and Supplementary Fig. 4). Many spin valves achieved an MR response of > 100%; and in **Fig. 2d** we show the performance characteristics for the spin valve with an MR of 298%. We calculated the MR from Eq. 1 (red dots **Fig. 2d**) and observed very little voltage dependence. The deviation observed with champion efficiencies of much greater than 20-50% is likely due to immaculate surface chemistries with minimal interfacial reactions in those select devices (as discussed later). Supplementary Fig. 5 shows the *I-V* characteristics of other high MR spin valves.

Because the MR is measured on a macroscopic spin valve we can also analyze the current-dependent junction magnetoconductance (MC), $\Delta G_{J(I)} = I(1/V_{M\uparrow} - 1/V_{M\downarrow})$ where $V_{M\uparrow}$ and $V_{M\downarrow}$ are the voltages for B+ and B- that give the same current.(*27*) Three notable features stand out in $\Delta G_{J(I)}$ vs. current plot (**Fig. 2e**): (1) At low currents the dependence is linear while at higher currents there is a slight deviation from linearity (green-line, **Fig. 2e** is a linear fit to the negative current data). The linear response regime at low currents is consistent with other measurements of chiral MR and demonstrates that these spin valves exhibits a linear dependence at low bias in a two-terminal device which is characteristic of CISS enabled spin valves (see Supplementary Fig. 6 for analysis on other spin valves), corroborating the collinear charge-to-spin generation process through the (*S*-MBA)$_2$PbI$_4$ layer ($SP_{CISS} \propto I_c$).(*27*) The small non-linearity at higher currents suggests that the tunneling barrier becomes current (or bias)-dependent. The reduction at higher currents may also be explained by the magnon excitation mechanism,(*28*) i.e., injected hot electrons at a higher voltage excite magnons and randomize the majority and minority states leading to a decrease of the spin polarization at the surface of the FM electrode. (2) The $\Delta G_{J(I)}$ response in the negative and positive current regime are symmetric. Supplementary Fig. 7 shows $\Delta G_{J(I)}$ when plotted as a function of the absolute current, the values for positive current and negative currents overlap indicating the electrical magnetochiral anisotropy (EMCA)(*29*) does not play a dominant role. (3) The plot does not go to zero $\Delta G_{J(I)}$ at zero current. The best fit of $\Delta G_{J(I)}$ to a straight line (green-trace, **Fig 2e**) gives an offset of 0.28 µS. These features suggests that the underlying CISS mechanism violates time-reversal symmetry, as previously discussed in the literature.(*30*) However, the symmetry of the *I-V* curves indicates that the mechanism is independent of the current direction.

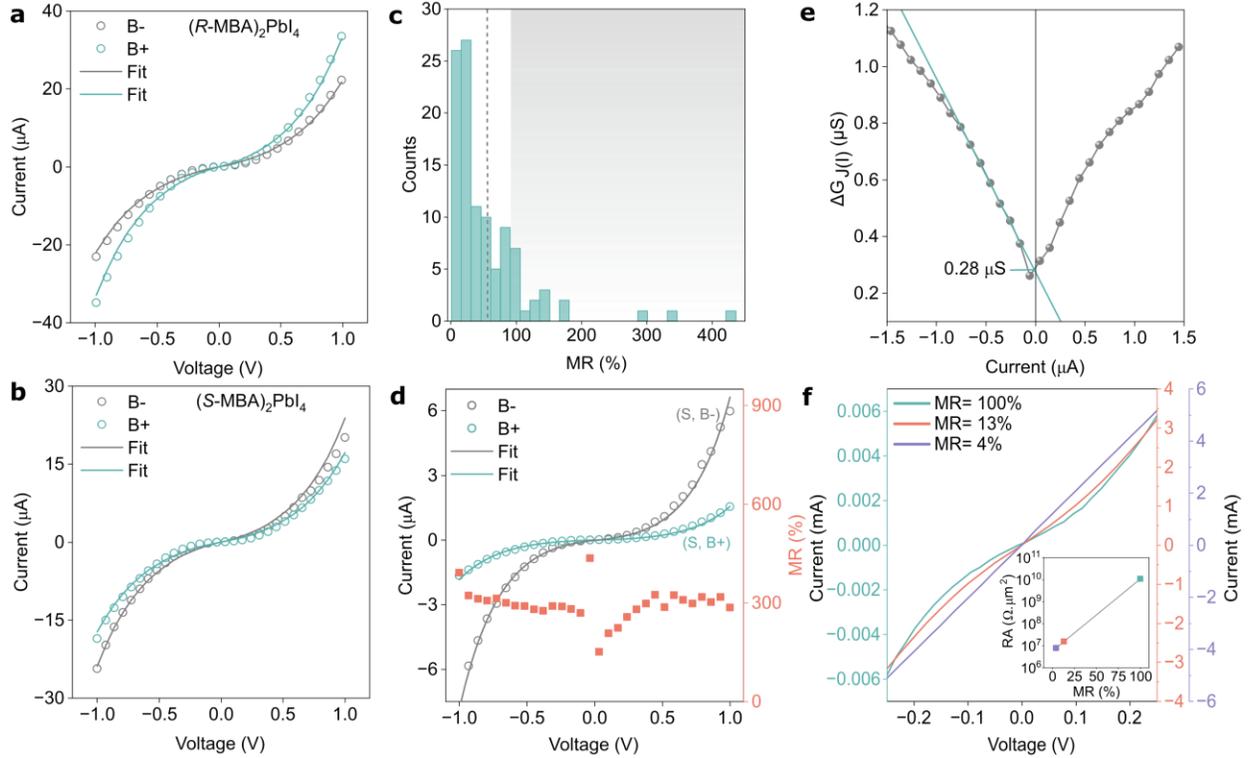

**Fig. 2| CMHS spin valves and conductance. a**, Representative *I-V* characteristics for (*R*-MBA)$_2$PbI$_4$ and **b**, (*S*-MBA)$_2$PbI$_4$ spin valves. Smooth lines are the best-fitted model discussed in the text (Eq. 6) to the data with only one adjustable parameter for both curves. Measurements were performed at magnetic field values of B±=±1T. **c**, Histogram of 106 spin valves for different CMHS compositions. Dotted line in c is indicative of the average MR value. The shaded region highlights the MR values of 100% and above. The curves in **a** and **b** are representative of the most common occurrence. **d**, *I-V* characteristics of 298% (*S*-MBA)$_2$PbI$_4$ spin valve from the histogram in **c**. **e**, $\Delta G_J$ determined from the data in **d** as a function of current showing a linear dependence at low currents (green-line is best fitted linear line) with a non-zero offset. **f**, Comparison of *I-Vs* under magnetic field ($B+ =+1T$) for spin valve (FM/Al$_2$O$_3$/(*R*-MBA)$_2$SnI$_4$/Au) with different strength of tunnel junction. Inset shows the RA product vs MR where MR decreases for leaky spin valves.

The interface between FM and metals or semiconductors play a critical role in determining the spin current injection efficiencies and resulting MR responses.(*31*) The MR values in the present work are significantly higher than previous reports, but consistent with high $g_c$ typically measured by mc-AFM.(*32, 33*) Since mc-AFM is performed at room temperature, even larger MR values are expected in optimized spin-valve architectures, where spin transport occurs across well-defined tunnel barriers. To elucidate the origin of the high MR, three spin valves with identical composition and architecture but differing MR magnitudes were analyzed. The variation in MR correlates with the effective tunneling behavior (non-linearity in *I-V*) of the junction, exhibiting a clear decrease in MR as the junction becomes more conductive.(*34-37*) Spin valves with leaky barriers show markedly reduced MR (**Fig. 2f**, Supplementary Fig. 8), underscoring the decisive role of tunnel barrier in governing spin transport. This can be better visualized as the resistance-area (RA) product defined as the product of the device resistance and its active junction area. In spin valves where a tunnel barrier is present, the effective barrier thickness and quality directly determine the RA value.(*38-40*) Here, we observed an increasing MR (inset Fig. 2f) with increasing RA, a feature observed in traditional tunnel junctions.(*41-45*) The discussion on the intricate balance between RA and MR is beyond the scope of the present work, however it should be noted that high RA is typical of spin valves employing insulating tunnel barriers.(*46-59*) RA for leaky spin valves is orders of magnitude lower and the corresponding MR approaches the low values generally observed in CMHS-based spin valves (Supplementary Fig. 9). Imperfections such as pinholes or incomplete barrier formation can facilitate spin-independent charge leakage, thereby suppressing spin selectivity and reducing the overall MR response.(*60*) These observations emphasize that achieving high-quality, well-defined interfaces is essential for realizing high CISS-MR.

Another interfacial effect that can also have strong implications in the context of the chiral semiconductors and the spin valves studied here involves interfacial magnetic proximity effects. It has been shown that the adsorption of chiral molecules on top of FMs can alter the remanent magnetization of the FM layer.(*61*) In addition, there are enantiospecific adsorption of chiral molecules on the FM surfaces depending on the magnetization direction of the FM.(*62*) These phenomena suggest that a strong chiral-modulated spin-exchange interactions occur between FM contacts and adjacent chiral molecules although no electron flow is present at the interface. This effect has been termed as magnetism-induced by proximity of adsorbed chiral molecules (MIPAC).(*61*) Although CMHS are substantially different than chiral molecules, we need to examine whether the MIPAC effect is also present at the FM/CMHS interface and if it plays a role in the observed MR responses. Thus, we fabricated macroscopic (*R/S*-MBA)$_2$PbI$_4$ spin valves without an Al$_2$O$_3$ tunneling barrier which allows for a direct contact between the FM and CMHS layer. Without the tunneling barrier, an ohmic *I-V* characteristic with high current levels is observed due to electrical short circuits between the top Au contact and bottom Ni contact. Even though the current does not transverse the CMHS layer we still observed an MR response, i.e., with the high conductance state for one enantiomer corresponding to a given magnetic field direction, and for the opposite enantiomer corresponding to the reversed magnetic field direction (Supplementary Note 2 and Supplementary Fig. 10-13). These results indicate a proximity effect between the CMHS and FM upon contact, even in the absence of direct current flow through the CMHS. We propose that this interaction arises from CISS, manifested through the MIPAC effect. The MR response from the MIPAC effect is much smaller compared to the CISS-MR measured in the spin valves with the tunnel barrier. The presence of the MIPAC effect indicates that there is a chirality-induced spin-exchange interaction between CMHS and FM similar to chiral organic molecules.

**CISS enabled spin-dependent tunnelling barrier**

The high values of MR and $g_c$ in the present case are in apparent contradiction with the Jullière model of magnetoresistance,(*63*) which should limit $g_c$ to be the product of the intrinsic spin polarization of the FM, $SP_{FM}$, and that of the CISS layer $SP_{CISS}$, i.e., $g_c = SP_{FM} \cdot SP_{CISS}$. Here the intrinsic spin polarization is defined as,

$$SP_i\% = \frac{n_{i,\uparrow} - n_{i,\downarrow}}{n_{i,\uparrow} + n_{i,\downarrow}} \times 100\%, \qquad [3]$$

where $n_{i,\uparrow}, n_{i,\downarrow}$ denote the density of states for spin up and spin down carriers $n$ at the Fermi level in each material $i$. As a result, in the Jullière model, the current dissymmetry fraction, $g_c$ should not be greater than the spin polarization of the FM, in our case, Ni with $SP_{FM}$ = 33% at room temperature.(64) Similarly, the MR is limited to be $MR = 2(SP_{FM} \cdot SP_{CISS})/(1 - SP_{FM} \cdot SP_{CISS})$ and thus should be limited to ~98.5% for Ni in the Jullière model assuming the maximum value $SP_{CISS} = 100\%$. However, the observed MR and $g_c$ greatly exceeded these limits in our experiments which suggests that $SP_{CISS}$ would be greater than 100%, which is not physical. This apparent paradox arises from oversimplifications in the Jullière model, which does not account for a spin-dependent tunneling barrier.(65-67) MacDonald and co-workers generalized the Jullière model to account for such a spin-dependent tunneling barrier.(68) In the MacDonald model, the intrinsic spin polarizations $SP_i$ of each electrode ($i$ = 1 or 2) in a spin-dependent tunnel junction should be each replaced by an *effective* spin polarization $SP_i^{eff}$ which can be written as,

$$SP_i^{eff} = \frac{f_i(\tau_i + 1) - 1}{f_i(\tau_i - 1) + 1}. \qquad [4]$$

In this model, the MR is given by,

$$MR = \frac{2\, SP_{FM}^{eff}\, SP_{CISS}^{eff}}{\left(1 - SP_{FM}^{eff}\, SP_{CISS}^{eff}\right)} \times 100\%. \qquad [5]$$

Here, the parameter $f_i = (SP_i + 1)/2$ where $SP_i$ is the spin polarization of electrode $i$, while $\tau_i = t_{i,\uparrow}/t_{i,\downarrow}$ is the ratio of the transmission amplitude factors for spin up vs spin down carriers in each electrode $i$; this assumes that the barrier tunnelling probability for each spin state can be factored as $\left|t_{\uparrow(\downarrow)}\right|^2 = t_{1,\uparrow(\downarrow)}\, t_{2,\uparrow(\downarrow)}$ (Supplementary Note 3). Assuming a symmetric tunneling barrier, $\tau_1 = \tau_2 = \tau$, the effective polarization of the FM is $SP_{FM}^{eff} = (f(\tau+1)-1)/(f(\tau-1)+1)$. As for the CISS layer, even assuming negligible intrinsic spin polarization, i.e., $SP_{CISS} = 0$, and $f_{CISS}$ = ½, the effective polarization of the CISS layer can be described as $SP_{CISS}^{eff} = (\tau - 1)/(\tau + 1)$. Thus, one can easily check if $\tau = 10$ and $SP_{FM} = 33\%$ then $g_c = SP_{FM}^{eff}\, SP_{CISS}^{eff} \times 100\% = 74\%$ and the corresponding $MR$ would be *570%*. For $\tau > 10$ the values of $g_c$ and *MR* could easily increase. For our observed MR value up to 298% (**Fig. 2d**), this model would imply that $\tau$ should be ~5.9 (see **Fig. 3** and discussion below).

To gain further insight into the spin-dependent tunneling and the calculated ratio of the transmission amplitude factors, $\tau$ for spin up vs spin down carriers in the CISS layer, it is useful to develop a predictive model that can account for the spin-dependent barrier physics and the true band structure at the interfaces. The non-linear *I-V* characteristics of the spin valves can be best described by a tunneling current through a rectangular barrier of height $E_b + \chi\Delta$ (**Fig.3,** Supplementary Note 3) where $\chi\Delta$ is the CISS-induced change in the work function is due to chiral-magneto interaction,

$$I \sim V \exp\left(-2s\sqrt{\frac{2m}{\hbar^2}(E_b + \chi\Delta - e|V|)}\right) \qquad [6]$$

with the applied bias $V$, the width of the tunneling barrier $s = 2$ nm, and the mass $m$ and charge of an electron $e$. Modeling the achiral *I-V* characteristics where $\Delta = 0$ yields a Al$_2$O$_3$ barrier height of $E_b$ = 2.9 - 3.3 eV consistent with literature.(*69*) Thus, for the CMHS spin valves $\Delta$ is the only adjustable parameter to describe both $I_{M\uparrow}$ and $I_{M\downarrow}$, and $\chi$ captures the symmetry of the *I-V* curves under reversal of chirality/magnetic field, where $\chi = +1$ represents the high resistance state and corresponds to the enantiomer and magnetization paring (R, -B) , and (S, +B), (**Fig. 3a,d**) while $\chi = -1$ is the low resistance state corresponding to the opposite pairings (R, +B) and (S,-B) (**Fig. 3b,c**). The experimental *I-V* characteristics are well reproduced by Eq. 6 consistent with our hypothesis (smooth lines in **Fig 2a,b,d**). Here we simultaneously model the +B and -B data to obtain the best-fitted value of $\Delta$. The CISS-induced change in work function found here varies between $\Delta = 20 - 100$ mV depending upon the exact CMHS. These values are of the same order of magnitude as found in specific measurements of the work function of peptide layers indicating similarities between CMHS and chiral molecules(*70-72*) and results in a spin-dependent tunnelling rate proportional to $\left|t_{\uparrow(\downarrow)}\right|^2$ as described above.

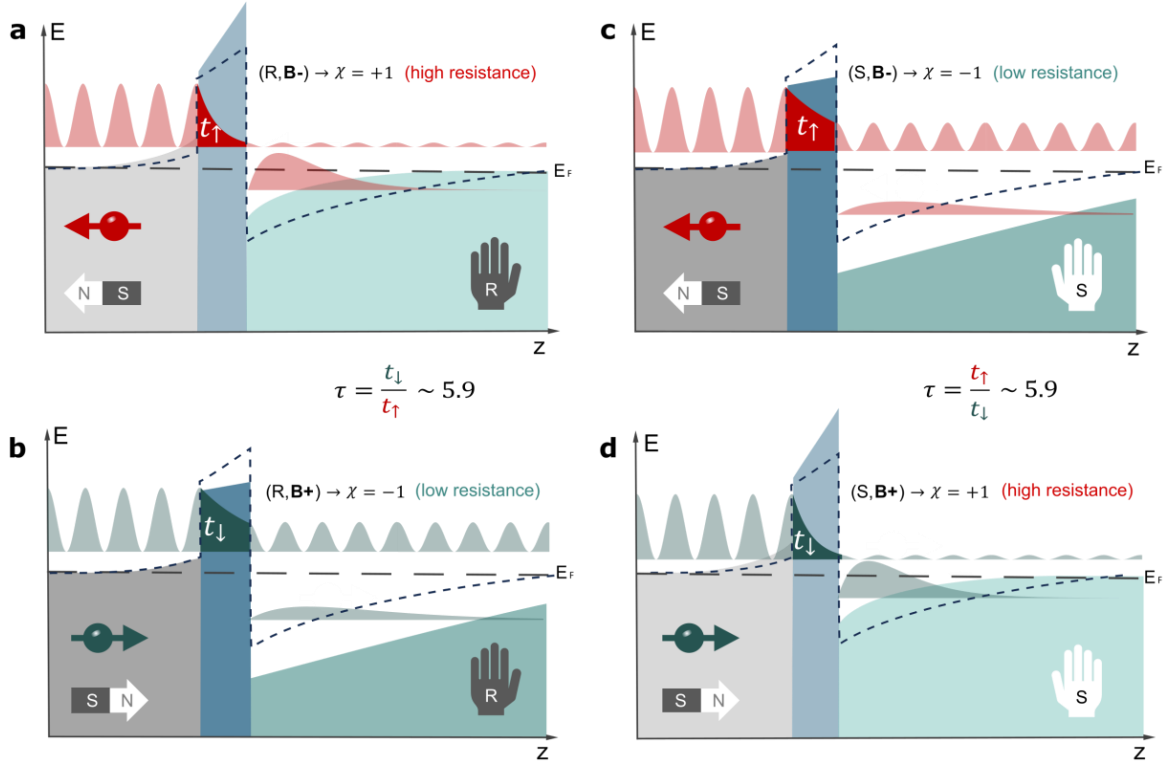

**Fig. 3| Modeling and CISS effect at the FM/CMHS interface. a-d**, Band alignment diagrams for the spin valves with tunnel barrier in the FM-dielectric-CMHS interface at zero bias voltage. The magnetic electrode is shown in gray, the tunnel electrode in blue, with the dashed line indicating the position of the tunnel barrier in an achiral state. Changes in the magneto-chiral local density of states affect the electrostatic screening at the interfaces and modulate the tunnel barrier. The chiral semiconductor corresponds to the green area. The four quadrants show all configurations of a spin valve and the symmetry of *I-V* curves, under reversal of chirality and magnetic field direction, with (**a,b**) and (**c,d**) showing opposite enantiomers (black and white hands) while (a,c) and (b,d) show opposite magnetization (denoted by the arrow). The diagonals (a,d) are in a high-resistance state with the symmetric pairing of (R,-B) and (S, +B), while (b,c) represent the low-resistance state with symmetric pairing of (R,+B) and (S,-B). When the magnetic field direction and chirality are matched the tunneling barrier is low and tunneling rate high. The ratio of the

tunneling rates, $\tau$, between the high resistance and low resistance state governs the MR discussed in the text and for the MR of 298% the ratio is ~5.9.

The CISS modulation of tunnel barriers has already been phenomenologically discussed elsewhere.(*73, 74*) In this context, it has been experimentally shown that a CISS modulation of the work function, decay distance, and quantum capacitance are observable consequences of the same effect without the need of charge transport across the chiral interface.(*72*) The square root in **Eq. 6** can be interpreted as a chirality and spin-dependent decay length $1/\lambda = \sqrt{\frac{2m}{\hbar^2}(E_b + \chi\Delta - e|V|)}$ or equivalently by a localization shift of the spin-dependent wavefunction leading to a chirality-induced quantum capacitance. However, the phenomenological energy Δ does not assume a specific mechanism of how the interaction with spin, charge, quasiparticles, and chirality causes this energy shift.

The effectiveness of this simple tunneling model arises from the spin valve design and operational parameters. Band edge alignment creates an ohmic contact between the top Au electrode and the CMHS, while band bending at the $Al_2O_3$/CMHS interface dominates the tunneling transport characteristics (**Fig. 3a-d**). The dielectric nature of $Al_2O_3$ modulates the tunneling barrier height based on effective work function differences. Since electron kinetic energy is well below the barrier, Fowler–Nordheim tunneling does not occur, making a rectangular tunnel barrier a valid approximation. The applied potential induces field changes much smaller than those from band bending, with chirality-induced effects being more significant. A back-of-the-envelope calculation shows a field change of 50 MV/m from chirality-induced effects, compared to less than 0.6 MV/m from a 0.1 V applied external bias (Supplementary Note 4). So, a small work-function change can induce significant changes in the transport characteristics. Notably, conductance differs at zero bias between opposite enantiomers under the same magnetic field polarity or the same enantiomers under opposite polarities. Thus, the model reproduces the chirality/magnetic field symmetry observed in the *I-V* characteristics and explains the deviation from the Jullière model due to the spin-dependent chirality-induced tunneling rate at the interface. However, to reiterate: the microscopic origin of the work-function modulation, captured in the model via the phenomenological energy Δ, especially its dependence on enantiomer and magnetization, remains unresolved although there are mechanisms proposed recently.(*75, 76*) The interaction of the chiral semiconductor with magnetic layer impacts the local density of states such that when the chirality and magnetic field are aligned (low resistance state, Fig. 3 b,c) there is less screening and lower barrier leading to a tunneling rate that is higher than when they are anti-aligned (high resistance state, Fig. 3 a,d). This phenomenology suggests a time-reversal symmetry breaking at equilibrium, independent of transport, and motivates further theoretical development beyond the scope of the present work.

**CISS-MR dependence on composition and symmetry**

Compared to traditional chiral systems, the synthetically tunable composition and high crystallinity of CMHS thin films allow both tailoring and quantification of the structural symmetry breaking in addition to the metal-halide composition, thus enables the ability to correlate the symmetry breaking to the resultant MR characteristics. For instance, the choice of chiral organic cation in the CMHS determines its space groups with distinct mirror symmetry breaking. The two most common space groups for CMHS are the $P2_1$ and $P2_12_12_1$ space groups.(*32*) The $P2_1$ space group has an in-plane $2_1$ screw axis even though the chiral organic cation's chiral axis is mainly oriented out-of-plane, while the $P2_12_12_1$ structures have $2_1$ axes along each crystallographic direction. Note that the $2_1$ screw axis symmetry operation by itself is not chiral, however, when chiral molecules are present in the structure they also lack mirror symmetry, in that case, the $2_1$ screw axis is also a chiral axis.(*77*) To elucidate this structure-property correlation, we selected four types of CMHS compositions with $P2_12_12_1$ and $P2_1$ space groups (**Fig. 4a**) by varying A-, B-, and X-sites,(*20, 78*) i.e., (*R/S*-MBA)$_2$PbI$_4$, (*R/S*-3BrMBA)$_2$PbI$_4$, (*R/S*-MBA)$_2$SnI$_4$, and (*R/S*-NEA)$_2$PbBr$_4$. We performed spin-dependent *I-V* characteristic measurements using the same architecture FM/Al$_2$O$_3$/CMHS/Au. Pure phase of CMHS thin films were obtained by dissolving their respective single crystals and spin-casting on the substrate (Supplementary Fig. 14). CMHS layer exhibits smooth and flat microstructure on the FM/Al$_2$O$_3$ substrate (Supplementary Fig. 15). **Fig. 4b** represents the MR calculated from *I-V* characteristics (Supplementary Fig. 16) of the macroscopic spin valves upon both positive and negative magnetic fields of 1T. Note that these values represent the most common MR results for each CMHS, i.e. the peak of the histogram in **Fig. 2c**.

We found that the CISS-MR follows the trend (*R*-MBA)$_2$SnI$_4$ > (*R*-MBA)$_2$PbI$_4$ > (*R*-NEA)$_2$PbBr$_4$ > (*R*-3BrMBA)$_2$PbI$_4$. We find the CISS-MR is *larger* for the $P2_12_12_1$ space group, where the $2_1$ screw axis aligns with the charge current direction, however a significant MR can be observed in the $P2_1$ space group despite the absence of collinear $2_1$ screw axis. Recent analysis of the spin textures for the $P2_1$ space group found symmetry-allowed chiral spin-splitting terms which produce spin polarization parallel to the carrier wave vector along both the out-of-plane and in-plane directions consistent with DFT simulations irrespective of whether a given direction coincides with the $2_1$ screw axis.(*79*) In fact, their work highlighted that these chiral spin-splitting terms arise due to the absence of any mirror symmetries rather than the presence of a specific screw axis, suggesting that neither the crystal space group nor the screw axis direction serves as a reliable measure *per se* to correlate with the CISS response.

To reveal the structural correlation, we compared two other metrics of chirality: circular dichroism (CD) and the continuous chirality measure (CCM). The CD dissymmetry factor $g_{CD}$ is an approximate empirical metric for the "degree" of chirality (in the absence of directionally asymmetric "apparent" CD effects which are uncorrelated to intrinsic chirality and can be observed in both chiral and achiral solids). CD spectra for all CMHS films exhibit an onset consistent with their optical absorption spectra (Supplementary Fig. 17-18), indicating successful chirality transfer to the inorganic framework.(*79*) We find that the $g_{CD}$ shows a trend when comparing CMHS within the same space group: (*R*-MBA)$_2$SnI$_4$ > (*R*-MBA)$_2$PbI$_4$ and (*R*-3BrMBA)$_2$PbI$_4$ > (*R*-NEA)$_2$PbBr$_4$ (Supplementary Fig. 19). The higher $g_{CD}$ in the case of (*R*-MBA)$_2$SnI$_4$ has been attributed to a higher mixing of the chiral A-site electronic states with Sn, compared to that in Pb, leading to the higher chiroptical activity.(*80*) The trend of $g_{CD}$ and MR is consistent for (*R*-MBA)$_2$SnI$_4$ and (*R*-MBA)$_2$PbI$_4$ (Supplementary Fig. 20).

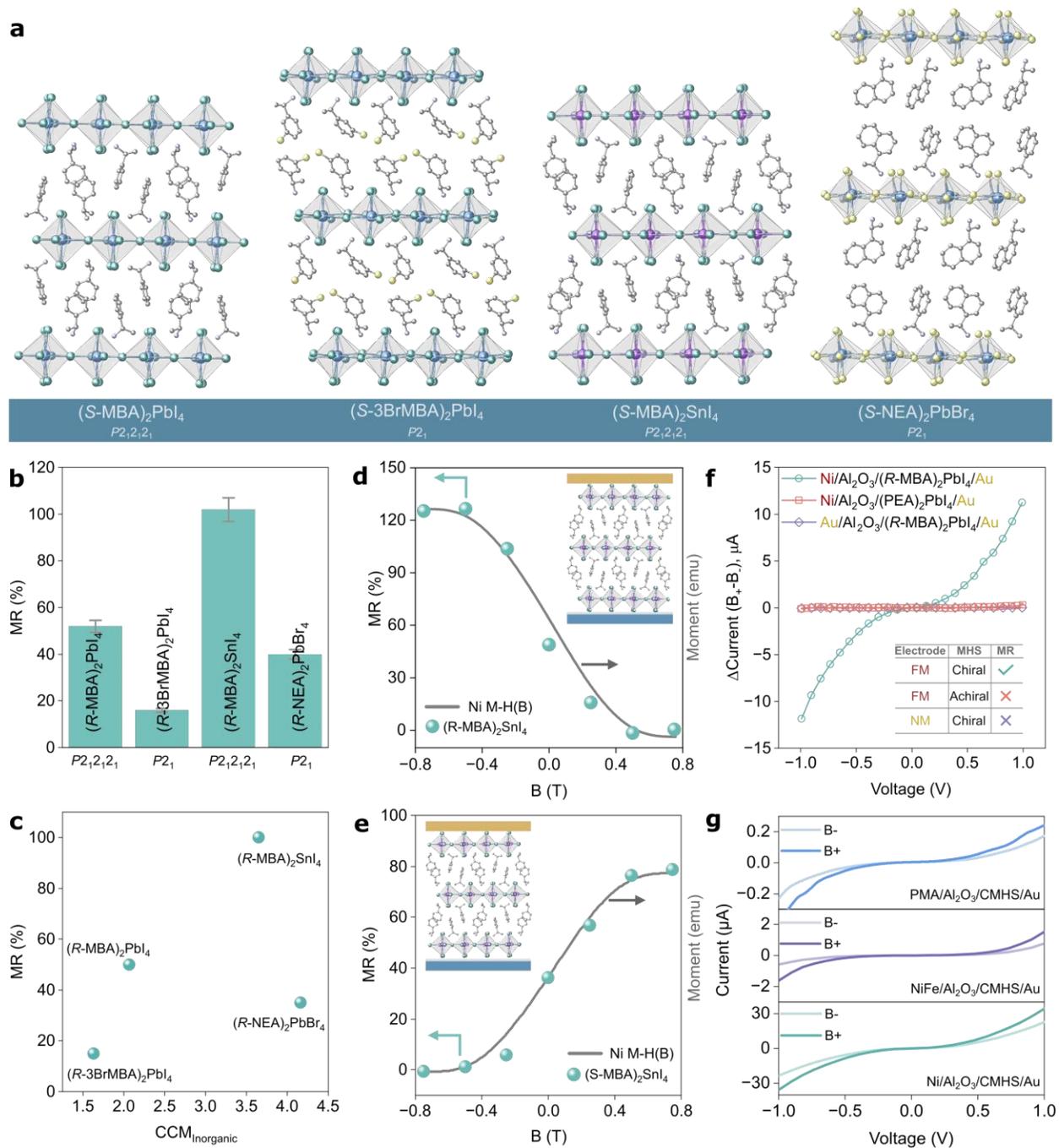

**Fig. 4| MR dependence on CMHS composition. a**, Crystal structure of (*S*-MBA)$_2$PbI$_4$, (*S*-3BrMBA)$_2$PbI$_4$, (*S*-MBA)$_2$SnI$_4$, and (*S*-NEA)$_2$PbBr$_4$. MBA-based CMHS have $P2_12_12_1$ symmetry while 3BrMBA- and NEA-based CMHS have $P2_1$ symmetry. **b**, Representative room-temperature MR values for various CMHS calculated from the current-voltage characteristics under out of plane magnetic field ($B\pm =\pm1$T). **c**, MR values as a function of inorganic CCM. Field-dependent MR(B) response for **d,** (*R*-MBA)$_2$SnI$_4$ and **e,** (*S*-MBA)$_2$SnI$_4$ spin valve at applied bias of -0.25 V. Green points are measured MR values by sweeping the field from positive to negative or vice versa. Gray lines show the magnetic hysteresis M-H(B) loop of the Ni layer measured by SQUID, which overlaps the MR(B) data, confirming the CISS-induced MR response in the current spin valve configuration. **f**, The difference in the current under B+ and B- magnetic field of $\pm1$T extracted from *I-V* measurements under different electrode and MHS configurations. MR response is only observed when both FM and CMHS are present. FM and NM denote ferromagnetic and nonmagnetic

contacts, respectively. **g,** Room-temperature *I-V* characteristics ($B\pm =\pm 1T$) for (*R*-MBA)$_2$PbI$_4$ spin valves with Ni, NiFe, and PMA (Pt/Co superlattices) FM layers. Fabrication details were similar for all spin valves. Ni and NiFe thickness were 30 nm and PMA layer consists of six alternating thin layers of Co and Pt.

On the other hand, CCM provides a quantitative descriptor for the degree of chirality (distortion) by calculating how far away the chiral structure is from the nearest achiral structure.(*81-83*) We used the racemic structure of each composition as a reference structure to calculate the CCM (Supplementary Note 5, Supplementary Fig. 21). We can also easily separate out the total CCM into the organic and the inorganic contribution. Interestingly they do not correlate with one another: Even though the organic component dominates the total CCM, the structures with larger organic CCM do not correspond to a larger inorganic CCM (Supplementary Fig. 22). There was no strong correlation when comparing $g_{CD}$ and MR as a function of CCM (**Fig. 4c**, Supplementary Fig. 23-24). To confirm that the spin valve response follows the typical magnetic polarization of the FM contact (Supplementary Fig. 25), field-dependent MR was measured. Field-dependent resistance measurements on (*R/S*-MBA)$_2$SnI$_4$ spin valves exhibit opposite MR response depending on the chirality of the CMHS layer and applied field (**Fig. 4d-e,** Supplementary Fig. 26). Further discussion on field dependence of MR and the protocol followed is provided in supplementary information (Supplementary Note 6 and Supplementary Fig. 27).

To validate that the resistance difference in the *I-V* characteristics truly originates from the CISS effect and simply not due to the presence of a magnetic field,(*84*) we fabricated control samples by replacing the FM contact with a non-magnetic electrode and CMHS with achiral MHS (PEA)$_2$PbI$_4$. The *I-V* response of the spin valves with the architecture Au/Al$_2$O$_3$/(*R*-MBA)$_2$PbI$_4$/Au and Ni/Al$_2$O$_3$/(PEA)$_2$PbI$_4$/Au show a negligible resistance difference under positive and negative fields (**Fig. 2f**, Supplementary Fig. 28-29), confirming that the FM contact and CMHS is essential to detect the CISS-MR response. To further demonstrate the impact of the magnetic field on the spin valve resistance, we measured *in-situ* continuous current response while reversing the magnetic field direction from -B to +B over several cycles (Supplementary Fig. 30). The current-time (*I-t*) response curve shows an abrupt change upon switching the field orientation, confirming the spin character of the CISS effect for the CMHS-based spin valve while the achiral MHS spin valve has no response to the magnetic field. To confirm the high MR observed is not specific to Ni, two other FM layers, NiFe and PMA (Co/Pt) were also employed in spin valves. NiFe and PMA-based spin valves show clear distinction between current values in opposite fields (**Fig. 4g**) confirming the CISS-induced MR in the spin valves. The lower current level in case of NiFe and PMA can be rationalized as the difference in the interface quality.(*85*)

**Conclusions**

The combination of macroscopic spin valve architecture exploration and measurements of multiple chiral halide perovskite compositions in this work enabled MR values to reach > 100 % in CISS-based spin valves. The correlation of MR with the CCM and orientation of chiral screw axes suggests that the CISS behaviors in chiral halide perovskites can be tailored and offer great potential for high throughput compositional screening. Furthermore, we revealed the presence of two types of spin-dependent interfacial effects: CISS and MIPAC based on the interfacial layer composition. We identified the important role of the spin-dependent tunneling barrier enabled by CISS in chiral spin valves which will provide device design rules in controlling spin at room temperature using solution-processed hybrid semiconductors. The molecular flexibility and tunable symmetry breaking of halide perovskites offer a compelling platform for the advancement of exploiting spin, charge, and light.

**Acknowledgments**
**Funding:** This project was supported by the Center for Hybrid Organic Inorganic Semiconductors for Energy (CHOISE) an Energy Frontier Research Center funded by the Office of Basic Energy Sciences, Office of Science within the U.S. Department of Energy. This work was authored in part by NLR for the U.S. Department of Energy (DOE) under Contract No. DE-AC36-08GO28308. The views expressed in the article do not necessarily represent the views of the DOE or the U.S. Government.

**Author contributions**
M.A.H., D.S., and M.C.B. conceived the idea and designed the experiments. M.A.H. performed all device fabrication and relevant electrical characterizations. I.A.L. and K.A. contributed to spin valve measurement in PPMS. P.M.T. and P.C.S. developed the theory. S.P.H. performed the TOF-SIMS measurements. J.T. helped with Ni layer deposition. M.P.H. synthesized the chiral perovskite single crystals. M.T. and Y.D. helped with optical characterizations. A.M. and A.H.C. contributed to FM layer fabrication and magnetic characterization. A.G. and Y.P. performed the CCM calculations. J.M.L. contributed to device design and analysis. M.A.H. and M.C.B. wrote the initial draft of the manuscript, and all authors contributed to the editing of the manuscript.

**Competing interests**
M.A.H., J.M.L., and M.C.B. are inventors on a provisional patent application related to the chiral perovskite-based spin valve reported here. Patent application number 63/900,252; country, USA. The other authors declare no competing interests.

**Data availability**
The data supporting the findings of this study are available from the corresponding author on reasonable request.

## Supplementary Materials for

Chirality-induced magnetoresistance in hybrid organic-inorganic perovskite semiconductors

Md Azimul Haque,[1,2] Pius Markus Theiler,[1] Ian A. Leahy,[1] Steven P. Harvey,[1] Jeiwan Tan,[1] Matthew P Hautzinger,[1] Margherita Taddei,[1] Aeron McConnell,[3] Andrew Greider,[4] Andrew H. Comstock,[3] Yifan Dong,[1] Kirstin Alberi,[1] Yuan Ping,[4,5,6] Peter C. Sercel,[7] Joseph M. Luther,[1,2] Dali Sun,[3] Matthew C. Beard[1,2]*

[1]National Laboratory of the Rockies, Golden, CO, USA
[2]Renewable and Sustainable Energy Institute, University of Colorado Boulder, Boulder, CO, USA
[3]Department of Physics and Organic and Carbon Electronics Lab (ORaCEL), North Carolina State University, Raleigh, NC, USA
[4]Department of Materials Science and Engineering, University of Wisconsin-Madison, Madison, WI, USA
[5]Department of Physics, University of Wisconsin-Madison, Madison, WI, USA
[6]Deparment of Chemistry, University of Wisconsin-Madison, Madison, WI, USA
[7]Center for Hybrid Organic Inorganic Semiconductors for Energy, Golden, CO, USA
*Corresponding author. Email: matt.beard@nrel.gov

**Methods**

**Chemicals and Materials**

R/S-α-methylbenzylamine (R/S-MBA), R/S-(1-naphthyl)ethylamine (R/S-NEA), $PbI_2$, $SnI_2$, $PbBr_2$, $SnO_2$, hypophosphorous acid, hydrobromic acid, N,N-dimethylformamide (DMF) were purchased from Sigma-Aldrich. Hydroiodic acid was purchased from Thermo Fisher Scientific. (R/S)-1-(3-Bromophenyl)ethanamine (R/S-3BrMBA) was purchased from synthonix. All reagents were used without further purification.

**Single crystal growth**

(R/S-MBA)$_2$PbI$_4$: Lead (II) iodide (413 mg, 0.895 mmol) was dissolved in 5.5 ml of concentrated hydroiodic acid and 0.5 ml of concentrated hypophosphorous acid. The solution was then cooled over ice and 228 μl (1.80 mmol) of R/S-MBA was added dropwise. The mixture was then heated at 120 °C until it was fully dissolved. The solution was then placed in a 75 °C aluminum heating block and cooled to room temperature at a rate of 1°C per hour. Orange needle-like crystals were collected through vacuum filtration, and the crystals were washed with 20 ml of diethyl ether three times followed by drying in a vacuum oven at 55 °C, 150 Torr.

(R/S-NEA)$_2$PbBr$_4$: $PbBr_2$ (90.0 mg, 0.245 mmol) was dissolved in 1.0 mL of concentrated hydrobromic acid by heating at 85°C. After cooling the solution to room temperature, 2.4 mL of DI water was slowly added to the mixture. Subsequently, the solution was cooled to 0°C, and R/S-NEA (87 μL, 0.48 mmol) was added dropwise. This mixture was then heated at 95°C while stirring for a minimum of 2 hours to ensure complete dissolution of precipitates. The solution was then left to cool to room temperature at a rate of 2°C per hour and crystals were collected over vacuum filtration.

(R/S-MBA)$_2$SnI$_4$: $SnO_2$ (135 mg, 0.896 mmol) was completely dissolved in 5.5 mL of hydroiodic acid and 0.5 mL of concentrated phosphoric acid by stirring and heating at 120°C overnight. The solution was cooled to room temperature. R/S-MBA (228 μL, 1.80 mmol) was added dropwise and a precipitate formed. The mixture was heated while stirring at 120 °C (< 10 min) until all solids were dissolved. The vial was subsequently transferred to an aluminum heating block (preheated at 90°C) and cooled to room temperature at a rate of 1°C/hour. In a nitrogen atmosphere, orange rod-like crystals were collected via vacuum filtration followed by washing with degassed diethyl ether ($N_2$ sparged ~10 min at 0°C). The crystals were then transferred to a vacuum oven at RT and dried overnight.

(R/S-3BrMBA)$_2$PbI$_4$: $PbI_2$ (231 mg, 0.5 mmol) was dissolved in 9.0 mL of concentrated hydroiodic acid by heating at 100°C (<5 min). The solution was cooled to room temperature and R/S-3BrMBA liquid (151 μL, 1 mmol) was added dropwise to the solution, and an orange precipitate formed. The mixture was heated at 120°C while stirring until completely dissolved (~ 1 h). The solution was then placed in an aluminum heating block preheated to 70°C and cooled to room temperature at a rate of 1°C/hour. Orange crystals were collected via vacuum filtration and washed with 50 ml diethyl ether three times and then allowed to dry at 55°C overnight in a vacuum oven.

**Spin valve fabrication**

The ferromagnetic layer (30 nm) of Ni or NiFe were deposited on cleaned ITO by e-beam evaporation. Spin valves with PMA layer were fabricated by depositing Co and Pt via radio frequency (RF) sputtering and process was repeated six times to build up the heterostructure. A 2 nm tunneling layer of $Al_2O_3$ was deposited on the top of the FM by atomic layer deposition. Atomic layer deposition was performed with pulsed trimethylaluminum and water at 200 °C to form 2 nm $Al_2O_3$. Prior to CMHS film deposition, the substrates were subjected to ozone for 1 minute. For CMHS thin films, single crystals were dissolved in DMF with a concentration of 200mg/ml. CMHS films were prepared by spin-coating 50 μl precursor solution on the substrate at 4000 rpm for 30 s, followed by annealing at 100 ˚C for 10 minutes in a glovebox. Depending on the device architecture, BCP(15 nm)/MoO$_x$(15 nm)/Au(80nm) or MoO$_x$(15 nm)/Au(80nm) or only Au(80 nm) was deposited by thermal evaporation on the top of the CMHS layer to complete the spin valve fabrication.

**Spin valve characterization**
Electrical characterization (*I-V* characteristics) of the freshly fabricated spin valves were carried out using a custom-built sample holder connected to a Keithley 2400, controlled by MATLAB code. Each substrate/sample consisted of 6 spin valves that were measured in sequence. The current/voltage compliance level for each composition of CMHS was determined by trial until the device failed under test conditions. The *I-V* characteristics were collected by placing the sample holder under a fixed magnetic field of ±1 T at room temperature using Lakeshore narrow gap fixed magnet. Magnetic field-dependent measurements for spin valves with tunneling layers were measured in air using an electromagnet from GMW associates. For spin valves without a tunneling layer, magnetic field-dependent measurements were performed in a physical properties measurement system (Quantum Design, Inc.). Magnetoconductance measurements were carried out by applying a DC bias to the sample and measuring the resulting current. All measurements were performed at room temperature and in dark unless otherwise stated.

**Material Characterization**
XRD measurements were performed using Rigaku Ultima IV with Cu Kα radiation at ambient temperature. SEM images were taken on a Hitachi S-4800 scanning electron microscope. UV-Vis absorption spectra were taken on a Cary 7000 spectrometer. CD measurements were carried out using Olis DSM 170 spectropolarimeter. An ION-TOF TOF-SIMS V spectrometer was used for depth profiling of MHS films. AFM images were acquired in amplitude-modulation mode using a Nanoscope Dimension 3100 system (Digital Instruments) equipped with Olympus AC160TS-R3 cantilevers. Magnetic hysteresis measurements for Ni were performed using a Quantum Design Magnetic Property Measurement System equipped with a Superconducting Quantum Interference Device (SQUID) magnetometer. The sample was mounted in a non-magnetic sample holder and centered in the detection coils to ensure accurate measurement. Magnetic moment was recorded as a function of applied magnetic field at a fixed temperature of 300 K, with the field swept between ±1 T. Data were corrected for the diamagnetic background of the sample holder.

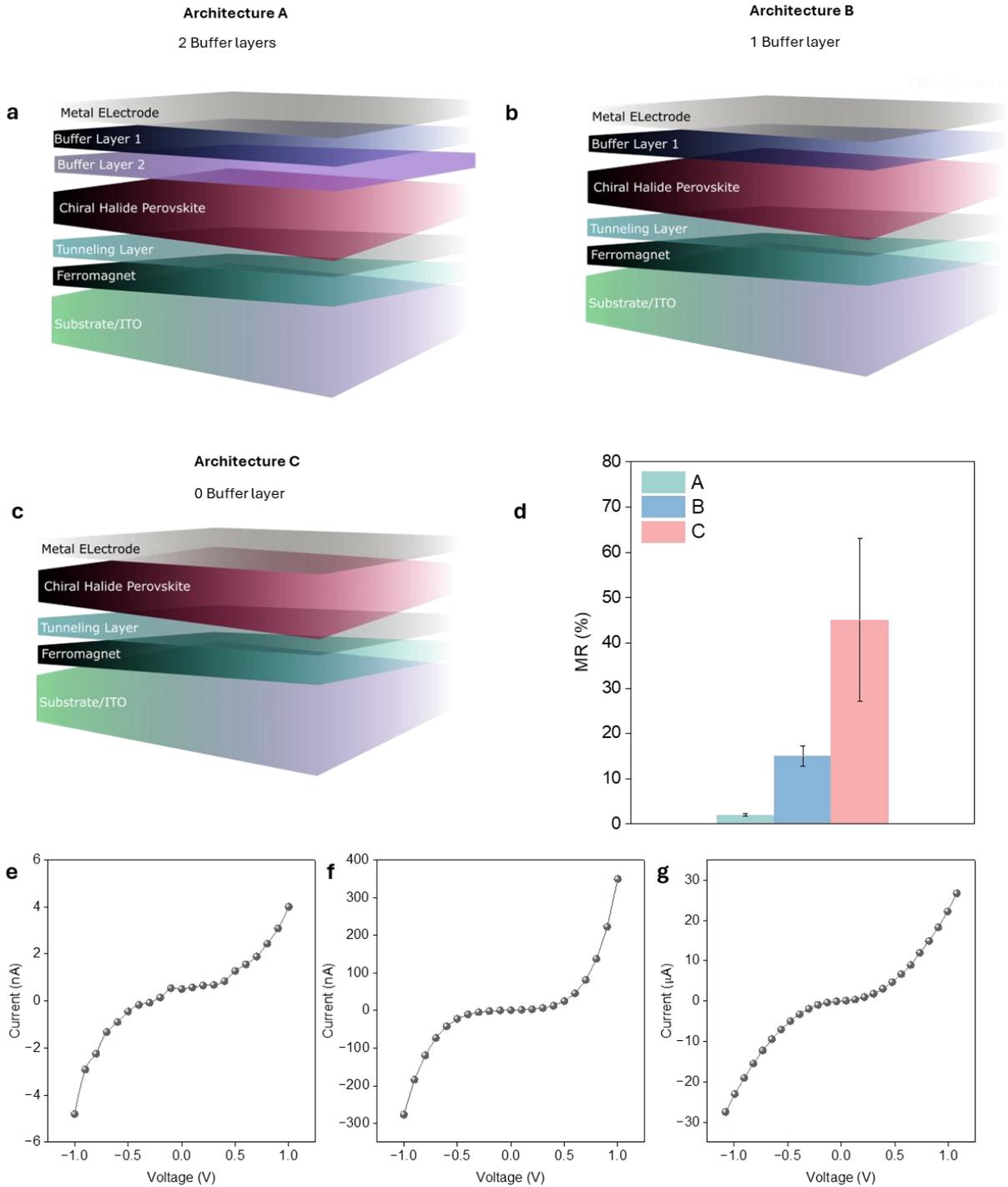

**Supplementary Figure 1.** CISS based spin valve architecture with **a** two buffer layers **b** a single buffer layer and **c** no buffer layer between the CMHS and metal electrode interface. **d** Average MR response as a function of spin valve architecture for (*R*-MBA)$_2$PbI$_4$. Buffer layer 1 and 2 represent MoO$_x$ and BCP, respectively. **e-g** *I-V* characteristics for spin valves with A, B, and C architecture, respectively. The current level increases with decreasing layer number.

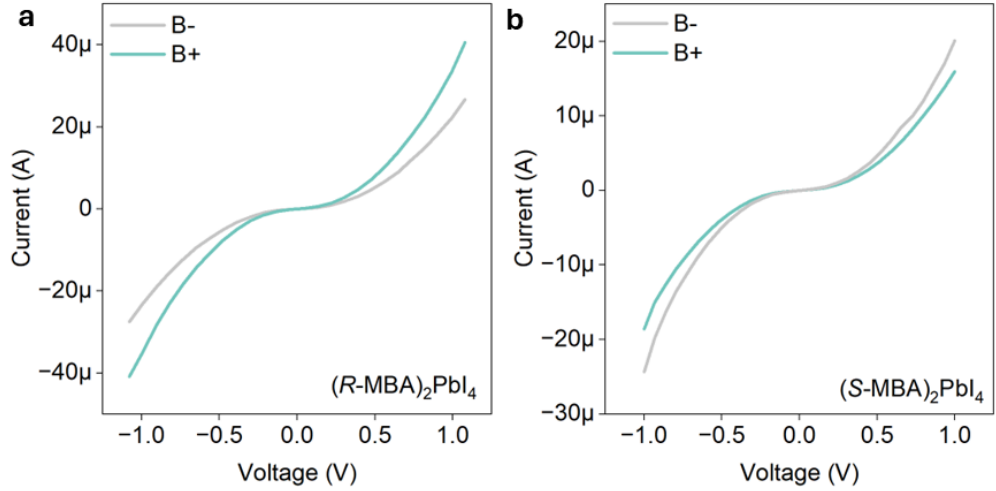

**Supplementary Figure 2.** Room-temperature current-voltage characteristics for **a** (*R*-MBA)$_2$PbI$_4$ and **b** (*S*-MBA)$_2$PbI$_4$ spin valves (FM/Al$_2$O$_3$/CMHS/Au) show opposite response under out of plane magnetic field (±1T) depending on the CMHS handedness.

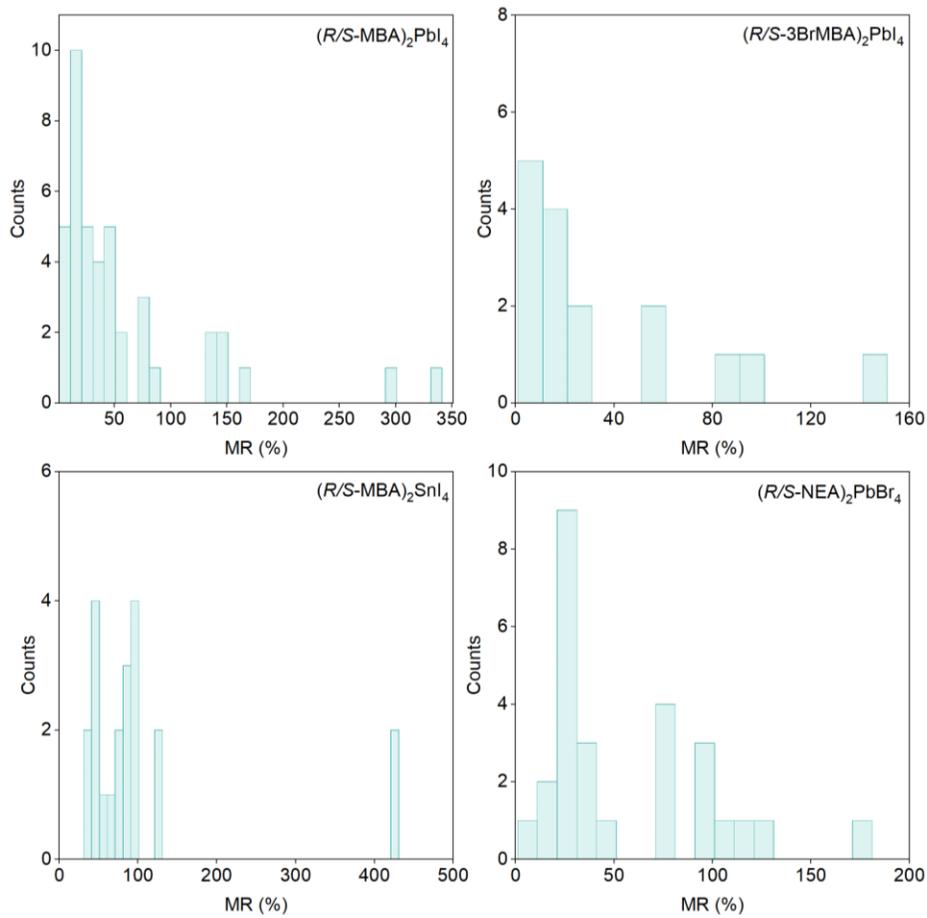

**Supplementary Figure 3.** MR histogram of spin valves for different CMHS compositions.

**Supplementary Note 1.** Spin valve fabrication guidelines

To ensure high quality spin valves, we follow a specific fabrication protocol. Supplementary Fig. 4 depicts the fabrication steps. ITO coated glass substrates were used to ensure good bottom electrode contacts. The substrates were masked to deposit ferromagnetic layers on the ITO. Afterwards the substrates were masked again to deposit the tunneling layer such that $Al_2O_3$ covers the edges of ferromagnetic layer to prevent electrical shorting with the top metal electrode. CMHS layer was coated on the tunneling layer and then CMHS layer was scratched off from the contact areas. Finally, top metal electrode was deposited by using a shadow mask using thermal evaporation. For electrical characterization, the compatible voltage range should be determined by scanning different voltage ranges (from low to high range). Starting the measurement from high voltage can break the spin valve. Spin valves with same architecture and composition can have variations in the performance. High MR spin valves have good tunnel junctions and MR decreases for leaky spin valves. The quality of CMHS layer synthesized from different batches of single crystals should be checked by CD measurements to ensure spin valve data reproducibility.

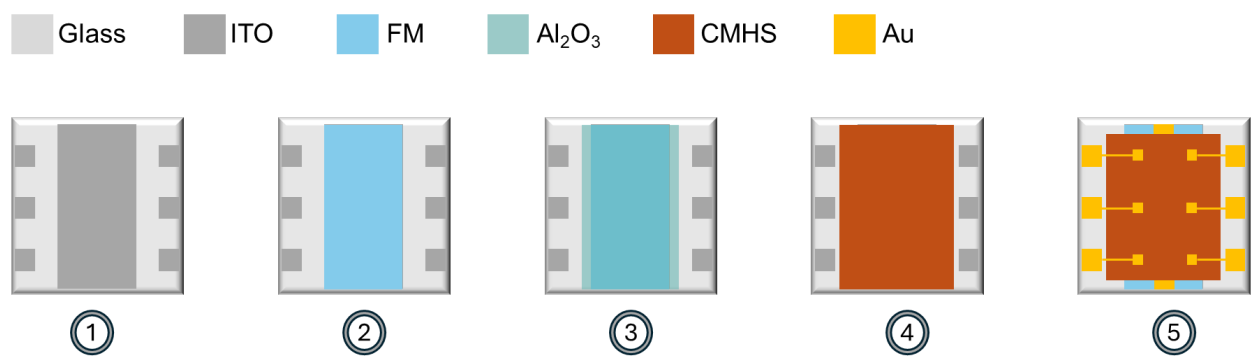

**Supplementary Figure 4.** Spin valve fabrication steps.

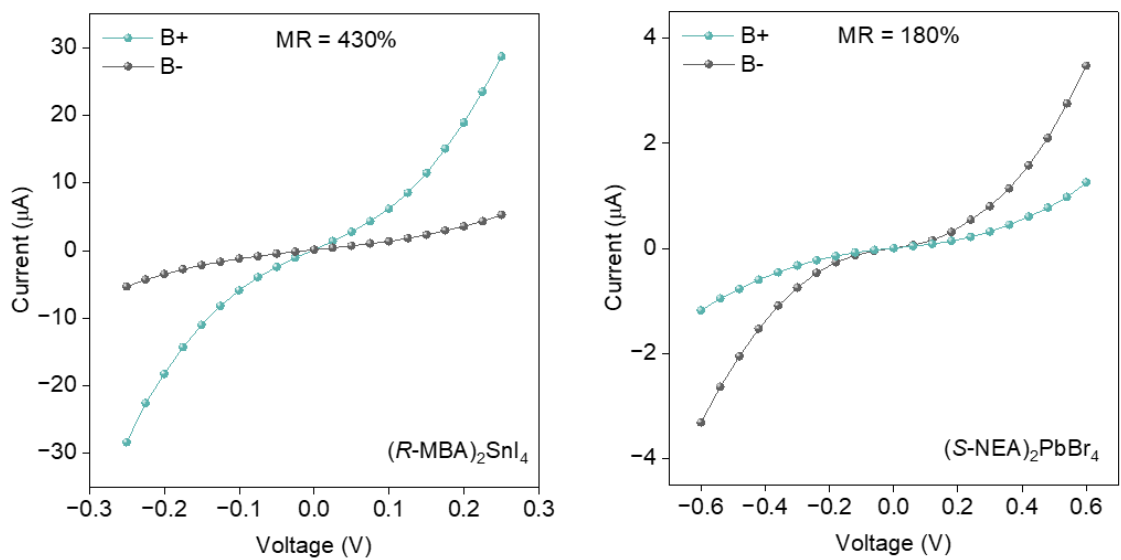

**Supplementary Figure 5.** I-V characteristics of high MR spin valves (FM/$Al_2O_3$/CMHS/Au) for (*R*-MBA)$_2$SnI$_4$ and (*S*-NEA)$_2$PbBr$_4$.

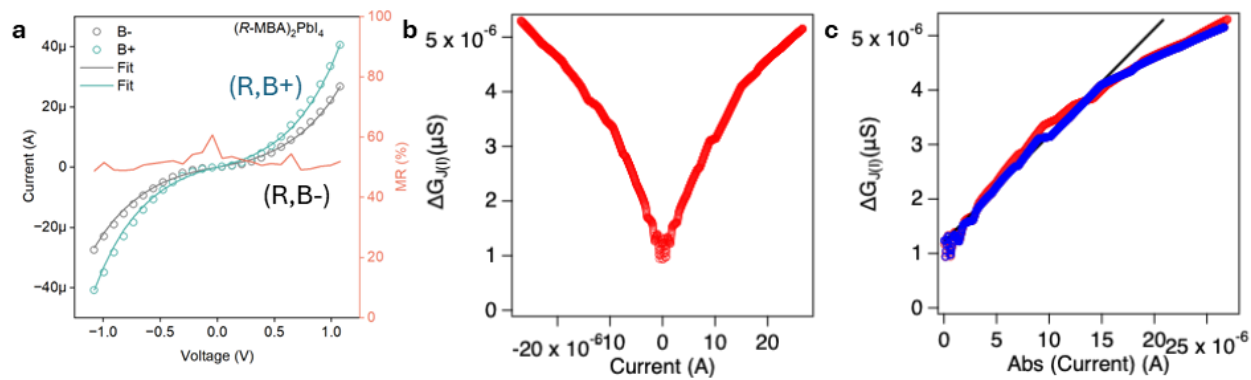

**Supplementary Figure 6. a** Fitting of the experimental data for (*R*-MBA)$_2$PbI$_4$-based spin valve (FM/Al$_2$O$_3$/CMHS/Au) by a model assuming tunneling current through a rectangular barrier. MR values calculated from *I-V* characteristics is also shown. $\Delta G_J$ as a function of current (**b**) and $\Delta G_J$ as a function of absolute current (**c**) for (*R*-MBA)$_2$PbI$_4$.

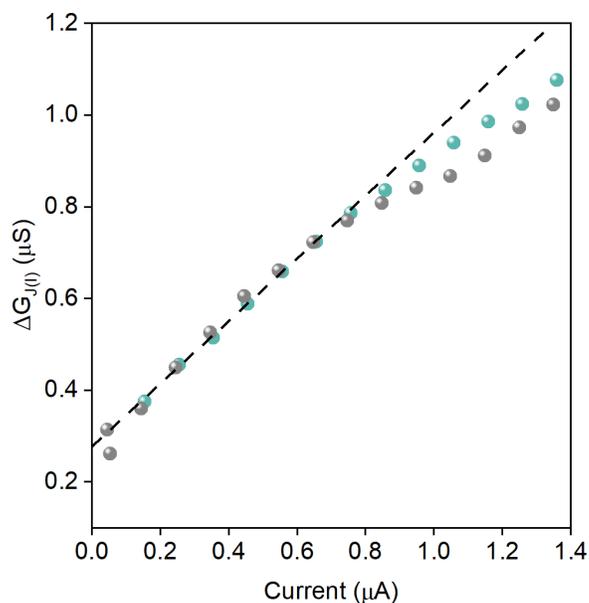

**Supplementary Figure 7.** $\Delta G_J$ as a function of absolute current for (*S*-MBA)$_2$PbI$_4$ spin valve (FM/Al$_2$O$_3$/CMHS/Au).

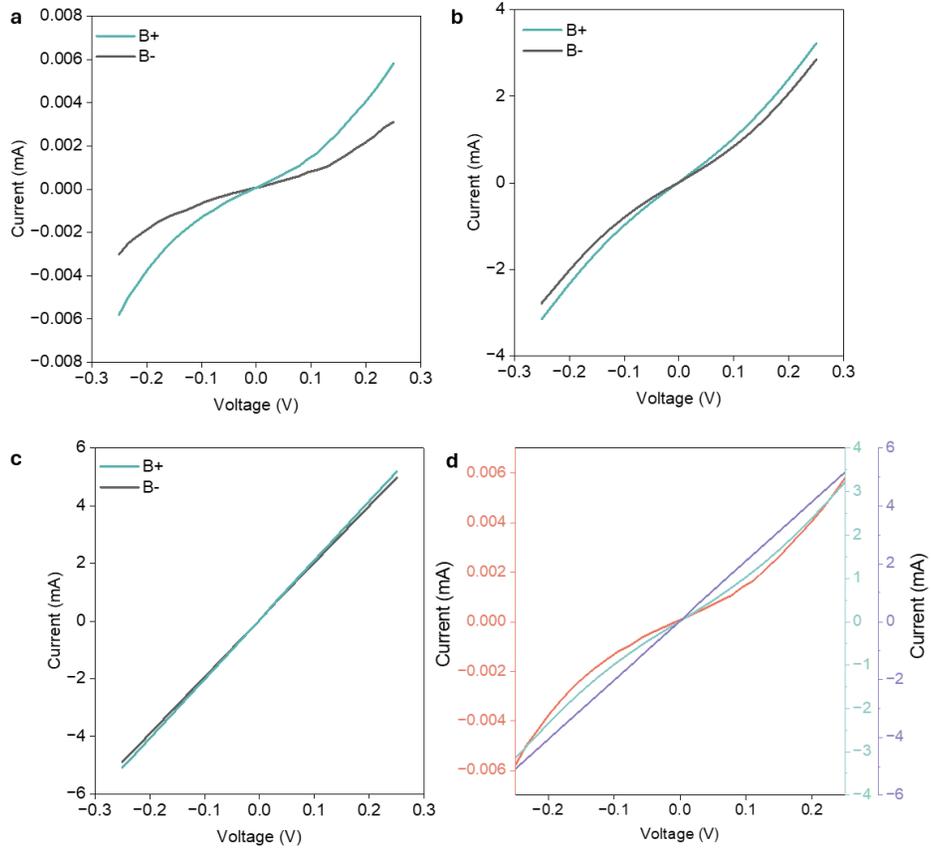

**Supplementary Figure 8. a-c** Current-voltage characteristics for (*R*-MBA)$_2$SnI$_4$ spin valves (FM/Al$_2$O$_3$/CMHS/Au) with different strength of tunnel junction. **d** Comparison of positive field I-V from a-c. MR decreases for leaky spin valves.

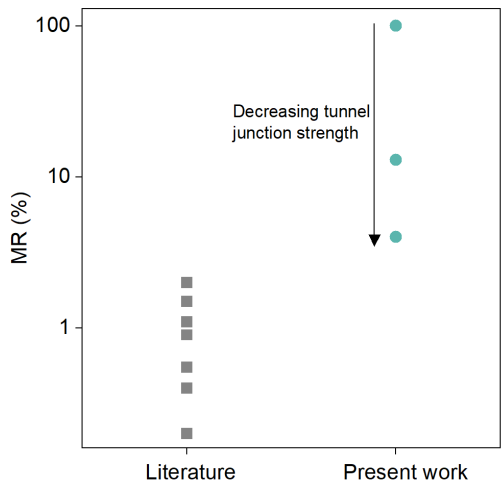

**Supplementary Figure 9.** Comparison of MR values for various CMHS from literature(*1-6*) with the present work for (*R*-MBA)$_2$SnI$_4$ spin valves with different degree of tunnel junction strength.

**Supplementary Note 2.** MR induced by magnetic proximity effect

Coupling between spin and charge transport arises at the interface of a FM and paramagnetic metal due to unequal spin-up and spin-down sub-bands in the FM.(*7*) Similarly, an interface between FM and antiferromagnet can create an exchange spring in the antiferromagnet due to the coupling to FM.(*8*) Furthermore, enhanced circularly polarized emission from FM/CMHS bilayers has been reported that was attributed to the formation of these chiral/FM interfaces, serving as another manifestation of the CISS effect.(*9*)

We fabricated macroscopic (*R/S*-MBA)$_2$PbI$_4$ spin valves without an Al$_2$O$_3$ tunneling barrier which allows for a direct contact between the FM and CMHS layer (Supplementary Fig. 10a-b). Without the tunneling barrier, an ohmic *I-V* characteristic is observed.(*10*) The current is in the mA range and not in the µA range observed from the corresponding spin valves containing the tunnel barrier. This larger current arises because of short circuits between the top Au contact and bottom Ni contact. The bottom Ni atoms can diffuse into the CMHS layer due to vacancies and chemical reactions adjacent to the metal/CMHS interface, resulting in short circuit. To confirm the formation of metal interdiffusion, the FM/CMHS interfaces were examined by depth-dependent elemental composition profiling using time-of-flight secondary ion mass spectrometry (TOF-SIMS). The presence of the tunneling barrier creates a sharp well-defined interface between the CMHS and FM layer. In contrast, the absence of tunnel barrier results in substantial interfacial mixing of CMHS and FM layers (Supplementary Fig. 10c, Supplementary Fig. 11).

For the *I-V* characteristics measured at $\pm$1T, the barrier free (*R*-MBA)$_2$PbI$_4$ device shows a higher and lower resistance under -B and +B, respectively (Supplementary Fig. 10d gray vs. green-traces). Conversely, when the handedness of the CMHS layer is inverted, the opposite trend is observed (Supplementary Fig. 10e). It is noteworthy here that there is still a MR response even though *no current* passes through the CMHS owing to the filament formation. We confirmed that a spin valve without having the CMHS layer, i.e., Au/Ni interface, does not show any difference in their *I-V* characteristics under +B and -B (Supplementary Fig. 12). In addition, we verified the short circuits in the device by measuring the anisotropic magnetoresistance (AMR) signals which is a fundamental characteristic of FM (Ni) layer. The MR(B) curve in the Ni/CMHS/Au device exhibits a characteristic butterfly-shaped curve, with a hysteresis dip around B = $\pm$0.1 T, corresponding to the coercive field of Ni (Supplementary Fig. 10f bottom panel). We found the CMHS interacts with the FM through the MIPAC effect revealing two key features: (1) the MR(B) loop exhibits asymmetry from -B to +B (Supplementary Fig. 11f top panel compare green vs gray at -1 T and +1 T) and (2) the coercive field of Ni decreases (dashed line in Fig 11f bottom to top panels compare bare Ni to CMHS/Ni). Supplementary Fig. 13 shows additional I-V characteristics of spin valves without the tunneling layer.

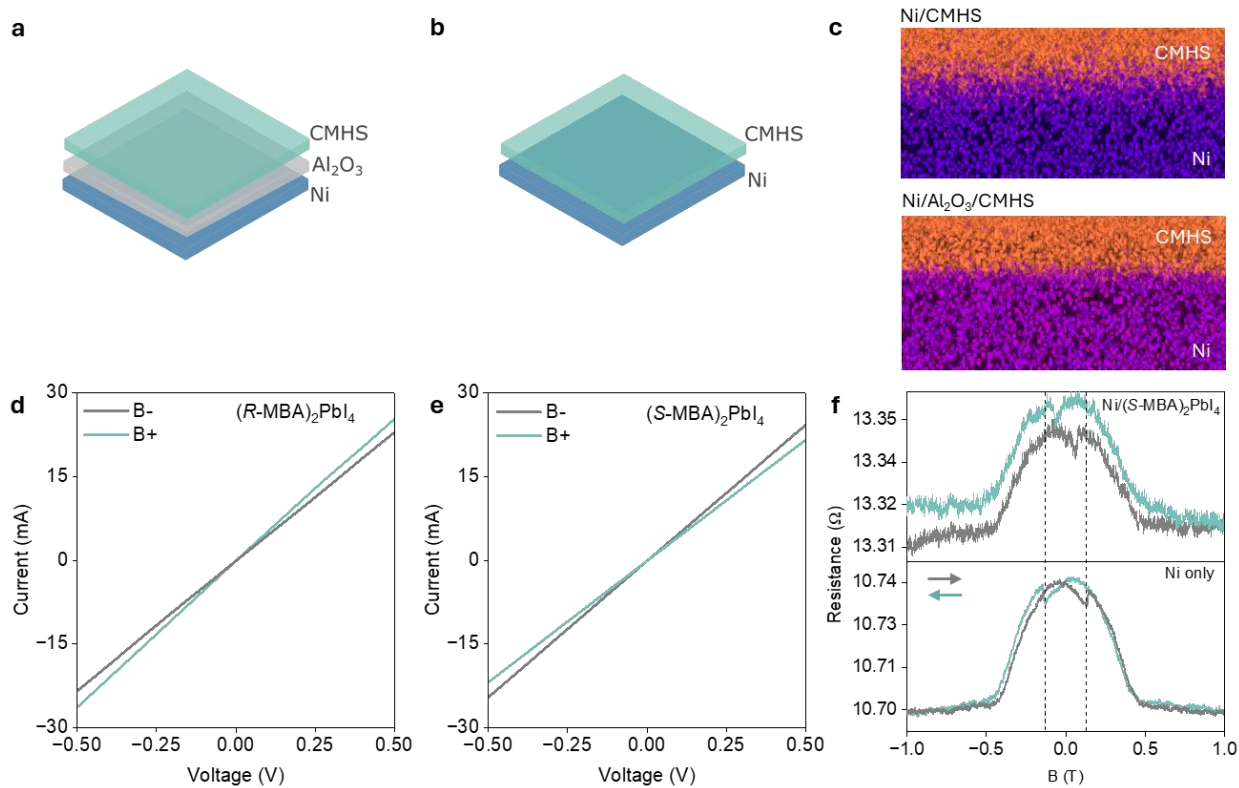

**Supplementary Figure 10.** Spin-polarized current exchange between CMHS and ferromagnet. **a,b**, Schematic showing the interface between the CMHS and the ferromagnet with and without $Al_2O_3$. For a typical CISS spin valve, there is an ultrathin tunneling layer present between the FM and active layer to mitigate conductivity mismatch. **c**, 3D rendering of TOF-SIMS for Ni/CMHS interface in the spin valve without and with $Al_2O_3$ layer. The absence of a tunneling layer leads to the mixing of Ni and Iodine while in the presence of $Al_2O_3$, a sharp interface is observed between Ni and CMHS. TOF-SIMS measurements were carried out after the electrical characterization of the spin valves. Current-voltage characteristics of tunneling layer-free spin valves (B±=±1T) of **d**, (R-MBA)$_2$PbI$_4$ and **e**, (S-MBA)$_2$PbI$_4$ CMHS. **f**, Magnetic-field-dependent response of spin valve without tunneling layer for Ni only and with (S-MBA)$_2$PbI$_4$ at -0.5V. I-V curves in (d-e) and magnetic field response (upper panel) in (f) are from different spin valves.

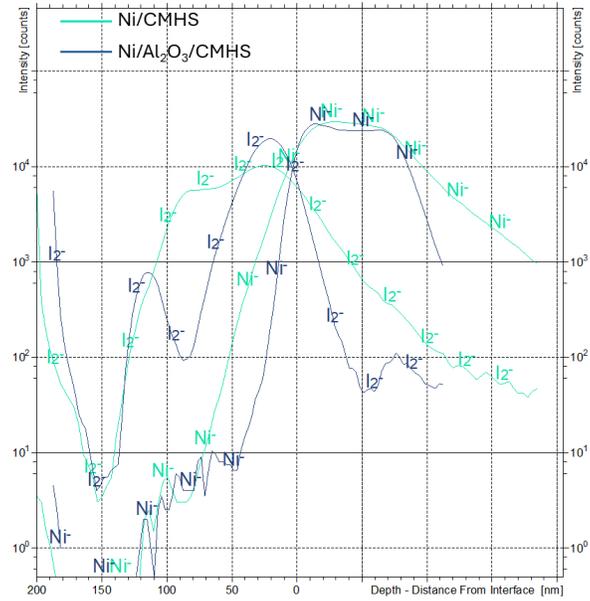

**Supplementary Figure 11.** TOF-SIMS depth profile Ni/CMHS interface in the spin valve without and with Al$_2$O$_3$ layer.

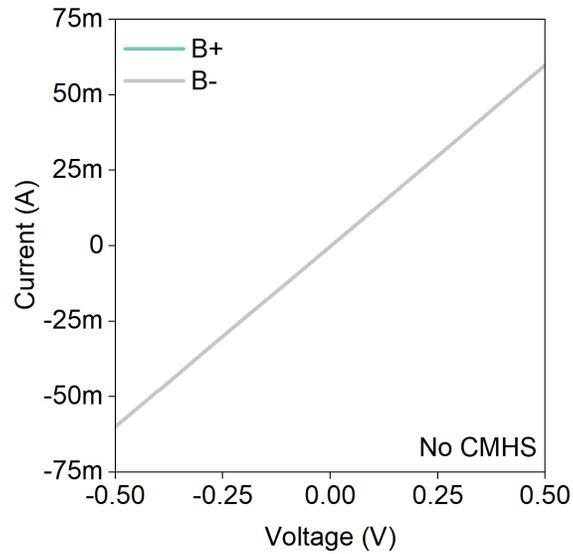

**Supplementary Figure 12.** Current-voltage characteristics of spin valves without CMHS.

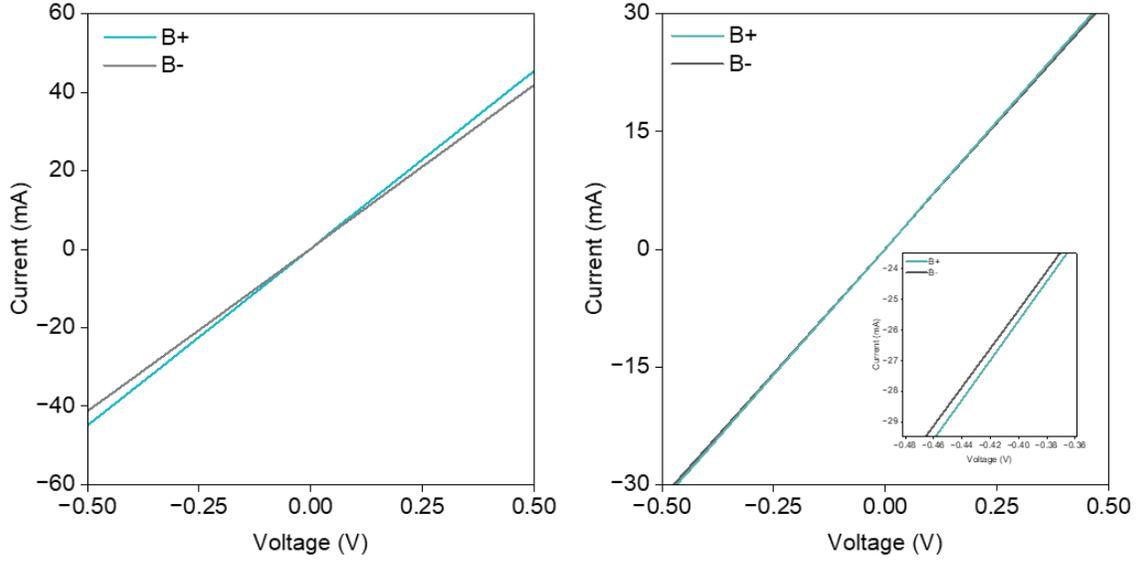

**Supplementary Figure 13.** Current-voltage characteristics of tunneling layer-free spin valves (B±=±1T) of (*R*-MBA)$_2$PbI$_4$.

**Supplementary Note 3.** Tunneling current through a rectangular barrier

The spin valve can be described in terms of a ferromagnet-oxide-semiconductor tunneling junction in the low bias regime.(*11*) The ferromagnet and the semiconductor are both connected to reservoirs with their corresponding chemical potential of $\mu_L = 0$ and $\mu_R = eV$. In the regime where the applied bias is much smaller than the tunneling barrier height $V \ll E_b$, the barrier can be approximated with a rectangular tunnelling barrier. Thus, we can write the Schrodinger equation of the system as:

$$\left[-\frac{\hbar^2}{2m}\frac{\partial^2}{\partial z^2} + U(z)\right]\Psi(z) = E\,\Psi(z)$$

With the potential landscape

$$U(z) = \begin{cases} 0 & z \leq 0 \\ E_b + \chi\Delta & 0 < z < s \\ 0 & s \leq z \end{cases}$$

where the chirality induced work function $\chi\Delta \ll E_b$. It depends upon the magnetization and the chirality. The wavefunction can be found using the ansatz $\Psi_j(z) = A_j \exp(ik_j z) + B_j \exp(-ik_j z)$ where $j = L, b, R$ together with requirement that the wave function is continuous as probability flux is conserved.

The wave vector for the barrier, the left and right contact are

$$\kappa = \sqrt{\frac{2m(E_b + \chi\Delta - E)}{\hbar^2}}$$

$$k = k_L = k_R = \sqrt{\frac{2mE}{\hbar^2}}$$

We are only interested in the transmission from left to right, where the electron from the left has an amplitude of $A_L = 1$, the reflected amplitude is $B_L = r$ and the transmitted amplitude is $A_R = t$ and $B_R = 0$ since no wave is coming from the right.

For the first interface,

$$\Psi_L(z=0) = \Psi_b(z=0) \quad \rightarrow \quad 1 + r = A_b + B_b$$

$$\left.\frac{\partial \Psi_L}{\partial z}\right|_{z=0} = \left.\frac{\partial \Psi_b}{\partial z}\right|_{z=0} \quad \rightarrow \quad k(1-r) = k_b(A_b - B_b)$$

and for the second interface:

$$\Psi_b(z=s) = \Psi_R(z=s) \quad \rightarrow \quad t\exp(iks) = A_b\exp(\kappa s) + B_b\exp(-\kappa s)$$

$$\left.\frac{\partial \Psi_b}{\partial z}\right|_{z=s} = \left.\frac{\partial \Psi_R}{\partial z}\right|_{z=s} \quad \rightarrow \quad k\,t\exp(iks) = i\kappa(A_b\exp(\kappa s) - B_b\exp(-\kappa s))$$

Solving for the transmission t, eliminating with some algebra yields the transmitted amplitude:

$$t = \frac{2\kappa k \exp(-2iks)}{2k\kappa \cosh(2\kappa s) + i(k^2 - \kappa^2)\sinh(\kappa s)}$$

The transmission probability is thus

$$T(E) = |t|^2 = \left(1 + \left(\frac{k^2 + \kappa^2}{2k\kappa}\right)^2 \sinh^2(\kappa s)\right)^{-1}$$

In the limit of the WKB approximation, at small energies ($E_b \gg E$) and wide barriers, $\kappa s \gg 1$ so that

$$T(E) \approx \exp(-2\kappa s)$$

Note that.

$$T(E) \approx \frac{4k\kappa}{(k^2 + \kappa^2)} \frac{4k\kappa}{(k^2 + \kappa^2)} \exp(-2\kappa s),$$

$$\approx t_1 t_2 \exp(-2\kappa s)$$

where the first and second terms $t_1 = t_2 = \frac{4k\kappa}{(k^2+\kappa^2)}$ respectively represent the tunneling amplitude from the left electrode into the tunnel barrier; and from tunnel barrier to the right electrode, while the exponential term describes decay in the barrier. Written in this way, it is clear that the factorization assumed in Eq. 3 of the main text is justified.(*11-14*)

Adapting Eq. 20 of Simmons,(15) the current through a tunneling barrier with similar electrodes is

$$I = \frac{Ae}{2\pi h\, s^2}\left[(E_b + \chi\Delta)\exp\left(-2s\sqrt{\frac{2m(E_b + \chi\Delta)}{\hbar^2}}\right) - (E_b + \chi\Delta - eV)\exp\left(-2s\sqrt{\frac{2m(E_b + \chi\Delta - e|V|)}{\hbar^2}}\right)\right]$$

which can be approximated for small, but non-zero voltages as

$$I \approx \frac{Ae^2 V}{4\pi^2\, \hbar\, s^2}\exp\left(-2s\sqrt{\frac{2m(E_b + \chi\Delta - e|V|)}{\hbar^2}}\right)$$

Since the exact surface area A of the junction is experimentally unknown, we can only describe the experimental proportionality

$$I \sim V \exp\left(-2s\sqrt{\frac{2m(E_b + \chi\Delta - e|V|)}{\hbar^2}}\right)$$

**Supplementary Note 4.** Electrical field strength estimation

The surface potential induced by chirality-related effects can vary by up to $\Delta U$ = 100 mV relative to the Fermi level of the ferromagnetic electrode. Assuming a tunneling barrier thickness of 2 nm and the absence of mobile charges within the barrier, the resulting electric field across it is given by

$$\left|\vec{E}\right| = \frac{\Delta U}{d} = \frac{0.1\,\text{V}}{2\,\text{nm}} \approx 50\,\text{MV/m}$$

In contrast, considering the entire device, which includes both the semiconductor and the tunneling barrier with a total thickness of 170 nm, the corresponding electric field is

$$\left|\vec{E}\right| = \frac{\Delta U}{d} \approx 0.6\,\text{MV/m}$$

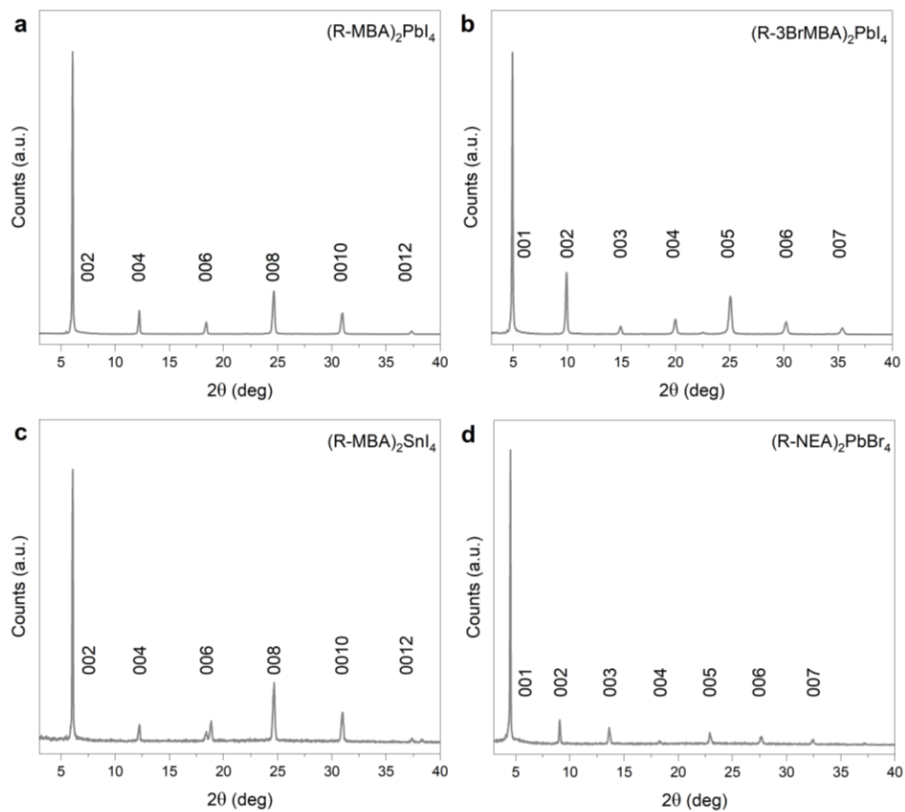

**Supplementary Figure 14.** XRD patterns of various CMHS films obtained by dissolving their respective single crystals in DMF and spin cast.

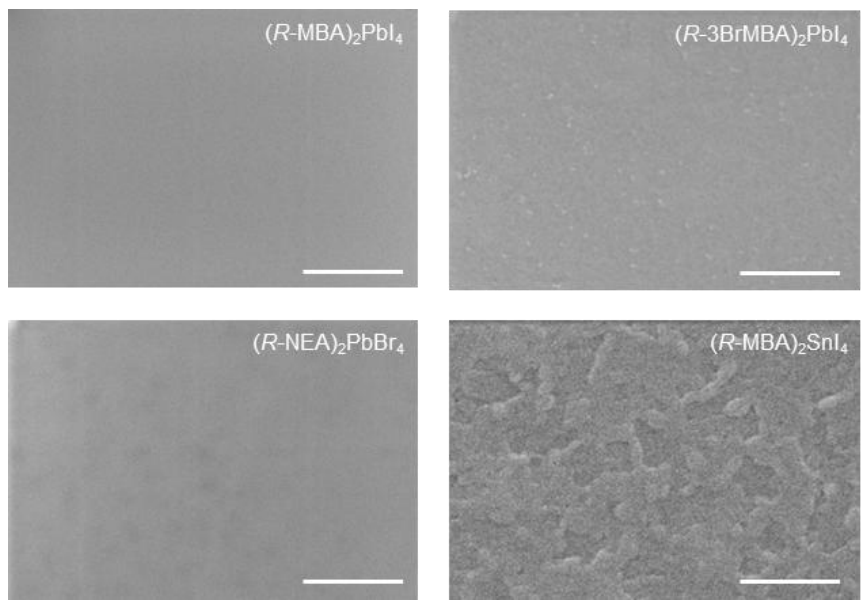

**Supplementary Figure 15.** Top surface SEM image of various CMHS layers coated on FM/Al$_2$O$_3$. The scale bar is 1μm.

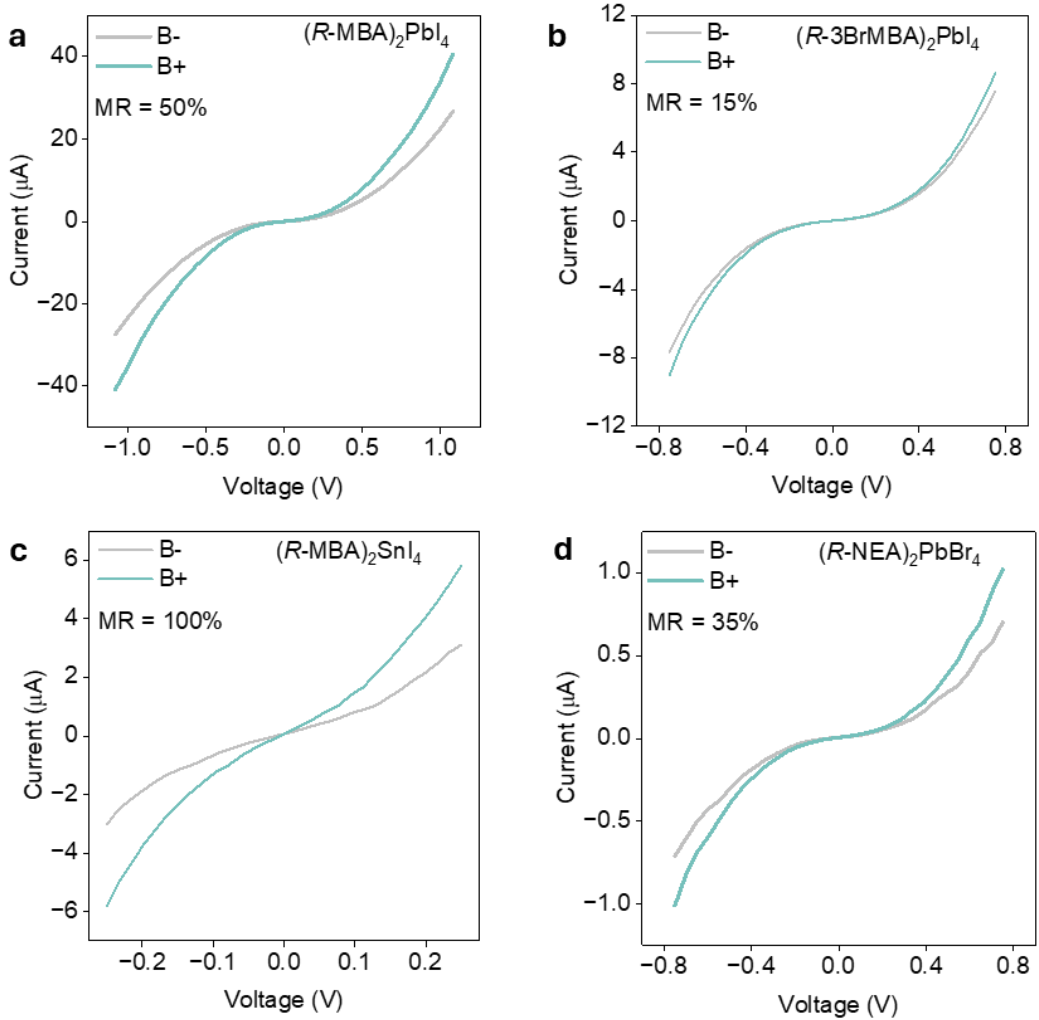

**Supplementary Figure 16.** Representative room-temperature current-voltage characteristics for spin valves (FM/Al$_2$O$_3$/CMHS/Au) employing CMHS. **b**, (*R*-MBA)$_2$PbI$_4$ **c**, (*R*-3BrMBA)$_2$PbI$_4$ **d**, (*R*-MBA)$_2$SnI$_4$ and **e**, (*R*-NEA)$_2$PbBr$_4$ under out of plane magnetic field ($B\pm$ =±1T).

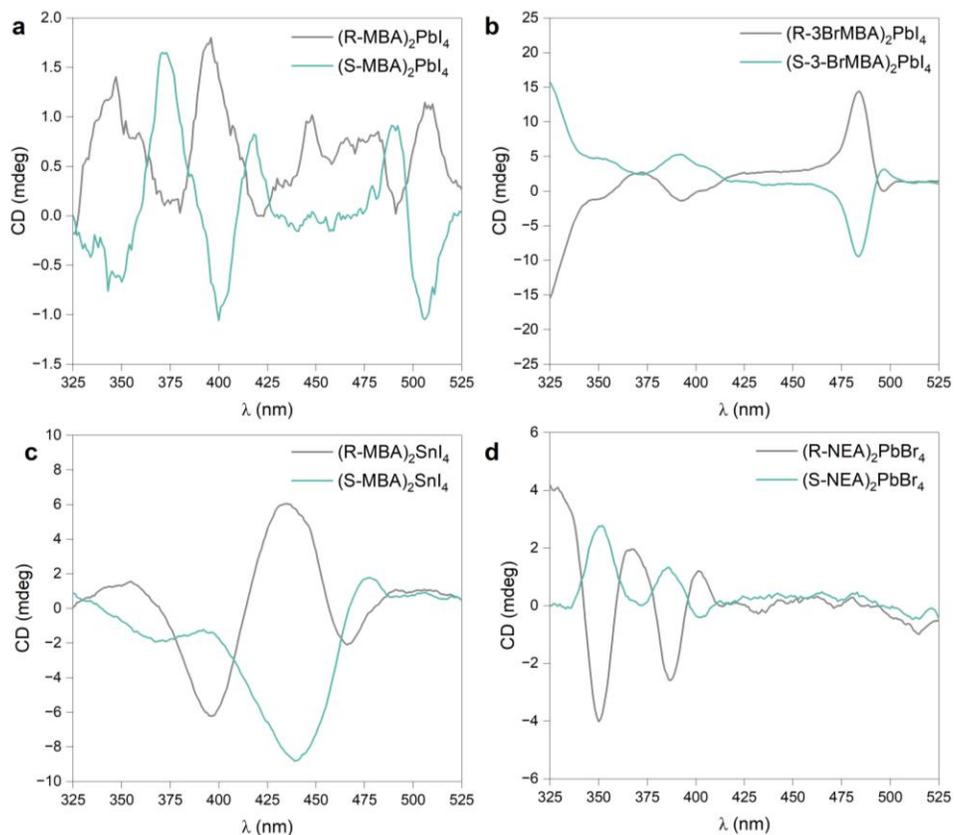

**Supplementary Figure 17.** CD spectra of various CMHSs. Films on glass were fabricated with the same recipe as that of spin valves.

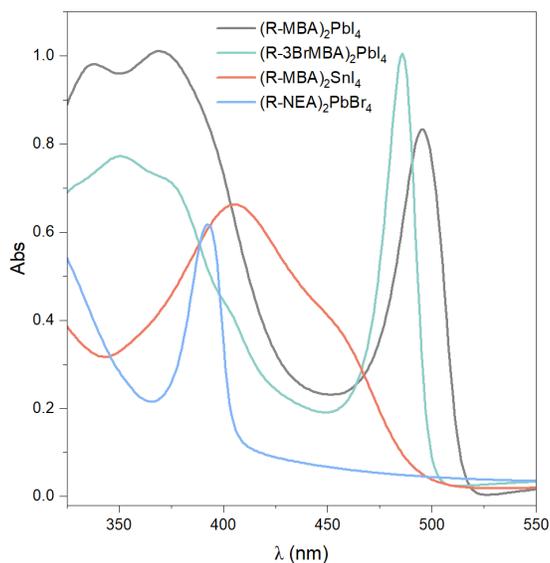

**Supplementary Figure 18.** Absorption spectra of various CMHSs. Films on glass were fabricated with the same recipe as that of spin valves.

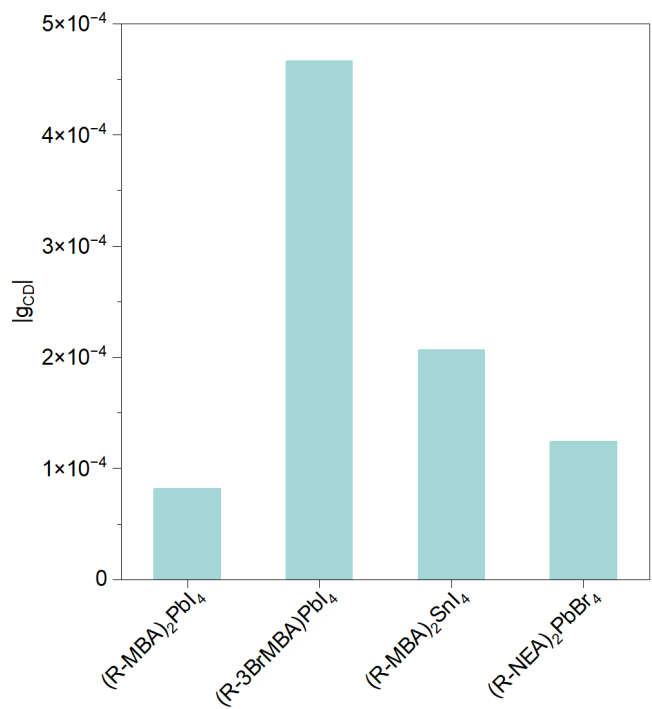

**Supplementary Figure 19.** Dissymmetry factor (g$_{CD}$) of various CMHSs calculated at the wavelength corresponding to the first major peak in the CD spectrum.

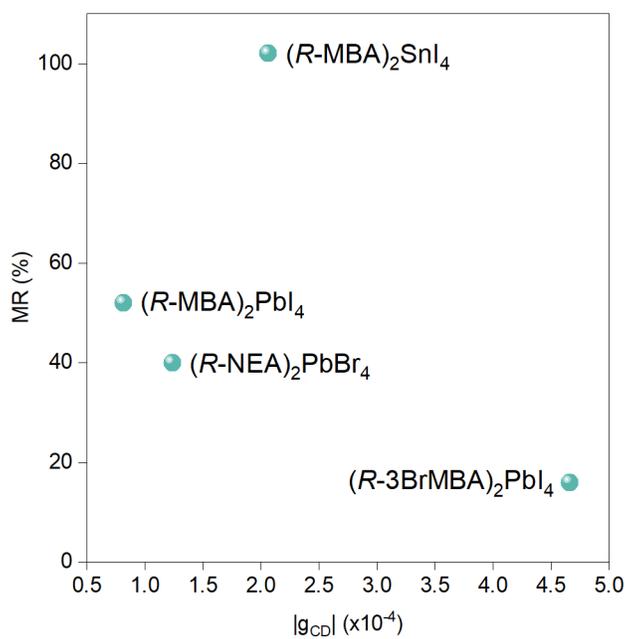

**Supplementary Figure 20.** Plot of $g_{CD}$ vs. MR for various CMHS.

**Supplementary Note 5.** Continuous chirality measure

Continuous Symmetry Measure (CSM)

The continuous symmetry measure (CSM) is computed by minimizing the difference of structure with atomic coordinates ($Q_k$) and the coordinates ($P_k$) belonging to point group symmetry ($G$). The set of points $P_k$ with point group symmetry $G$ are systematically varied until $\sum_k^N |Q_k - P_k|^2$ is minimized. The minimized difference is then normalized by the center of positions vector ($Q_0$). For the inorganic sublattice the CCM is computed using the racemic inorganic positions as $P_k$. The CCM is computed by minimizing the CSM, $S(S_n)$ with respect to the achiral point group ($S_n$).(16, 17)

$$CCM = \min[S(S_n)]$$

$$S(G) = 100 \cdot \frac{\min[\sum_k^N |Q_k - P_k|^2]}{\sum_k |Q_k - Q_0|^2}$$

$$Q_0 = \frac{1}{N} \sum_k^N |Q_k|^2$$

We find that the chiral molecules tend to have larger CCM than the inorganic layers and does not alone predict the amount of chirality in the inorganic layer. For instance, 3BrMBA has the largest CCM but introduces the least chirality into the inorganic layer. Additionally, MBA introduces more chirality into the Sn based perovskite than the Pb counterpart.

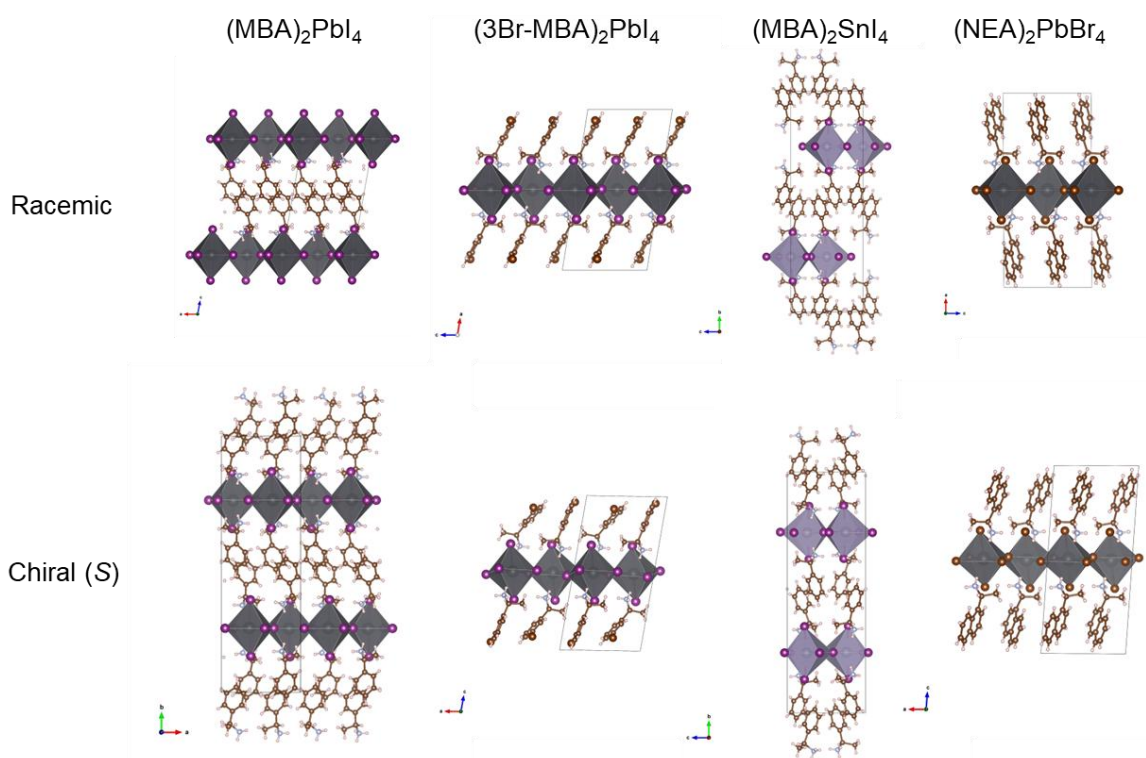

**Supplementary Figure 21.** Racemic and chiral crystal structures of various CMHS compositions used for CCM calculation.

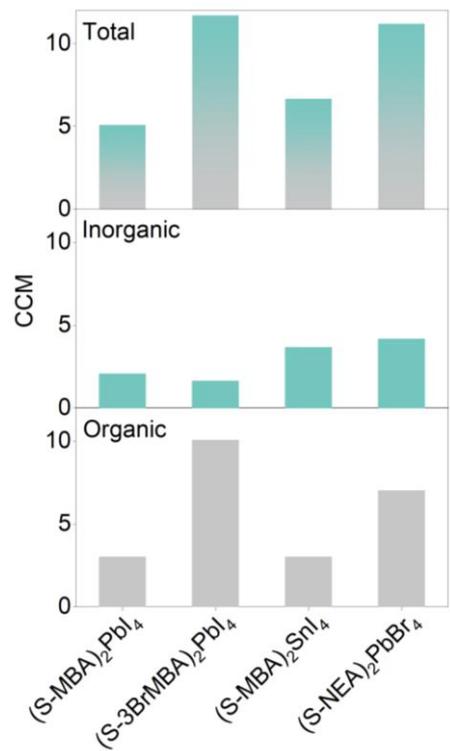

**Supplementary Figure 22.** CCM for the organic cation, inorganic framework along with total CCM for various CMHS compositions.

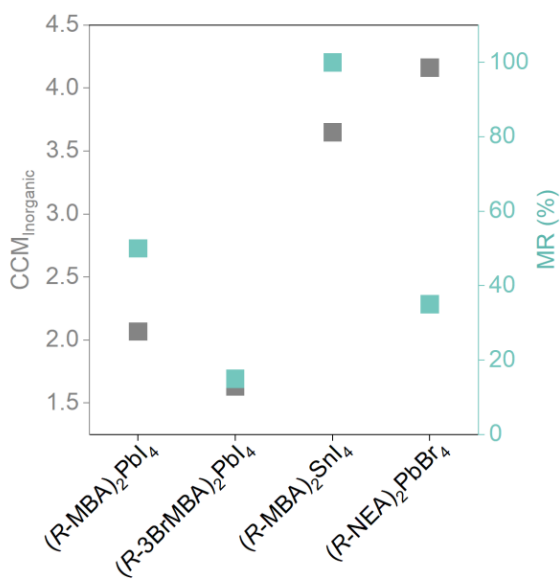

**Supplementary Figure 23.** Comparison between inorganic CCM and MR for various CMHS compositions.

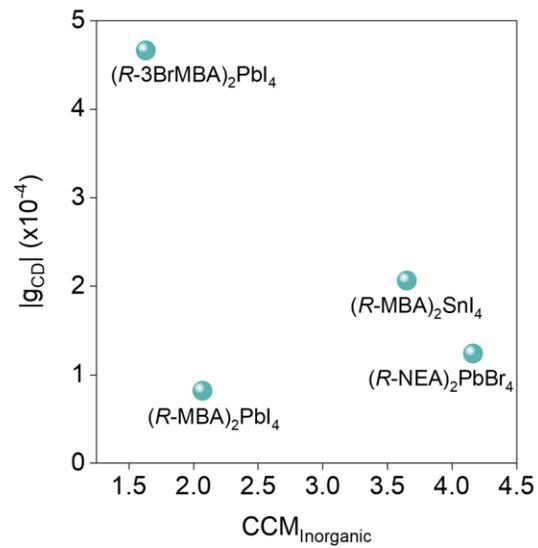

**Supplementary Figure 24.** Plot of $g_{CD}$ vs. CCM of inorganic lattice.

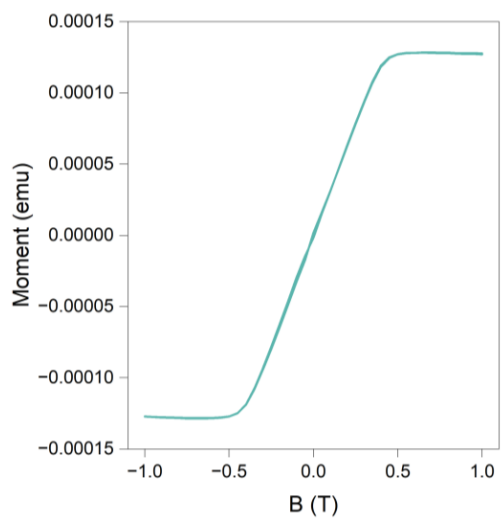

**Supplementary Figure 25.** Hysteresis of the FM (Ni) electrode.

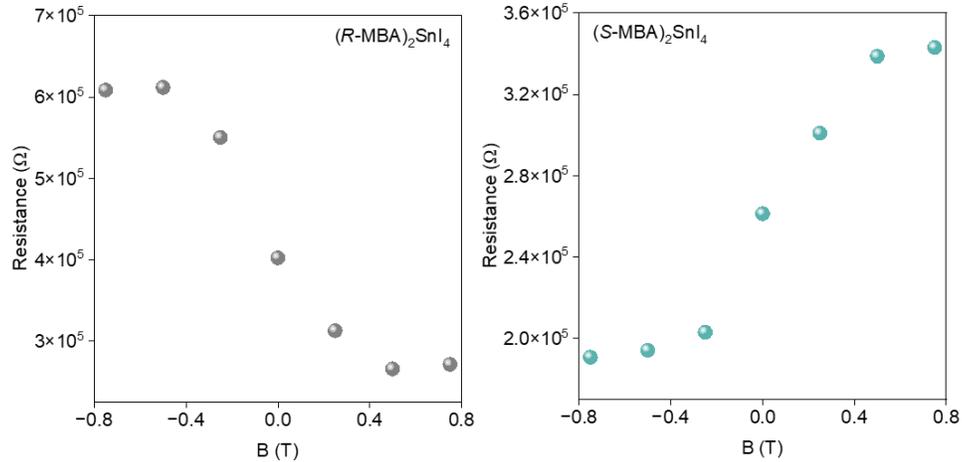

**Supplementary Figure 26.** Field-dependent resistance for (R/S-MBA)$_2$SnI$_4$ at -0.25 V.

**Supplementary Note 6.** Field-dependent MR

The electronic stability of CMHS needs to be considered for field-dependent measurements as current drift or device degradation can occur under prolonged electrical bias. To minimize these effects, resistance was measured only at a limited set of magnetic field values to construct the field-dependent MR curve. Because Ni electrodes exhibit negligible magnetic hysteresis, all measurements were performed while sweeping the field in a single direction, which provides sufficient data points to reproduce the expected Ni magnetization profile. For field dependent measurements, robust spin valves with negligible current drift and clear difference between multiple cycles of positive and negative fields were selected. An example of such a spin valve is shown in Supplementary Fig. 27. It is important to note that CMHS-based spin valves can become electronically unstable when subjected to repeated magnetic field sweeps, likely due to ion migration or interfacial degradation. To avoid electrical overstressing, resistance measurements should be performed using minimized dwell times and controlled field sweep rates. Developing optimized MR measurement protocols specifically tailored for CMHS devices will therefore be essential to achieve reliable and reproducible field-dependent magnetoresistance in future studies.

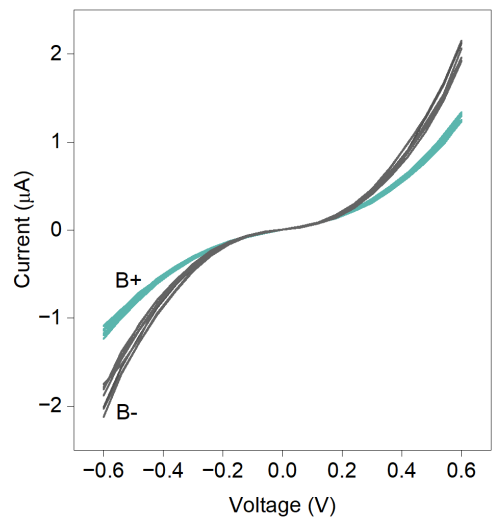

**Supplementary Figure 27.** Multiple cycles of I-Vs for spin valve (FM/Al$_2$O$_3$/CMHS/Au) employing (S-NEA)$_2$PbBr$_4$ CMHS under out of plane magnetic field (B± =±1T).

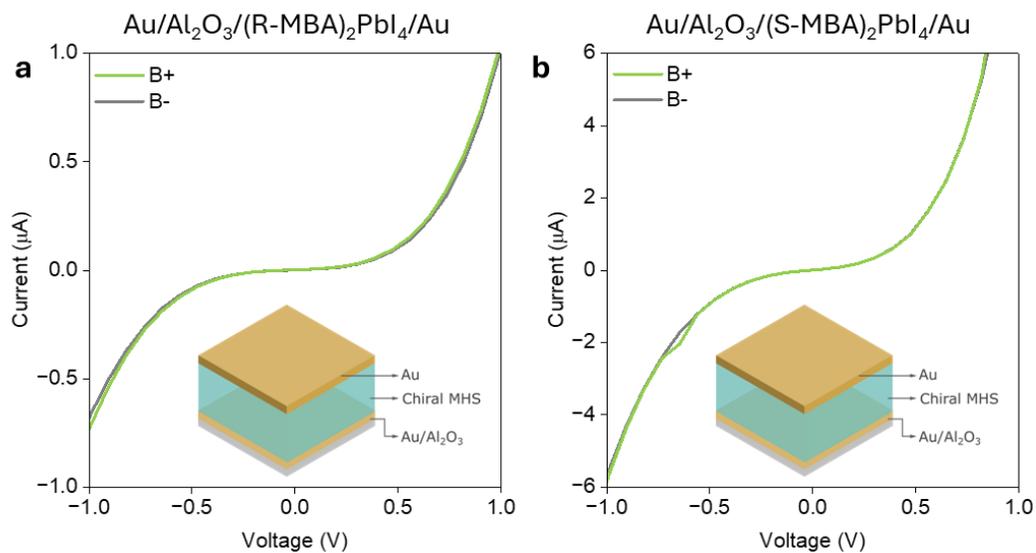

**Supplementary Figure 28.** Current-voltage characteristics of **a** Au/Al$_2$O$_3$/(R-MBA)$_2$PbI$_4$/Au **b** Au/Al$_2$O$_3$/(S-MBA)$_2$PbI$_4$/Au showing negligible difference between opposite magnetic fields.

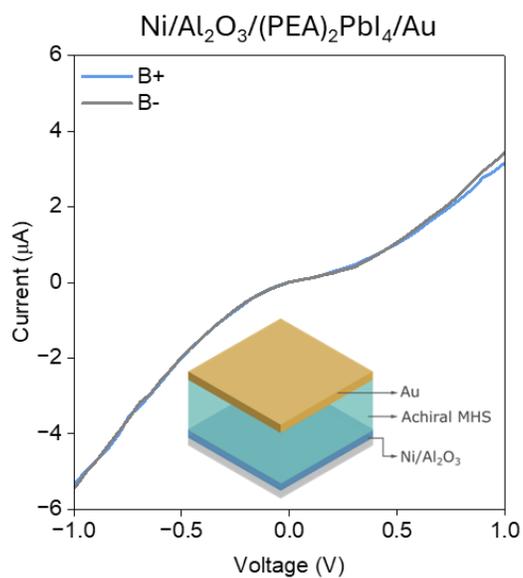

**Supplementary Figure 29.** Current-voltage characteristics of Ni/Al$_2$O$_3$/(PEA)$_2$PbI$_4$/Au spin valve showing negligible response to magnetic field.

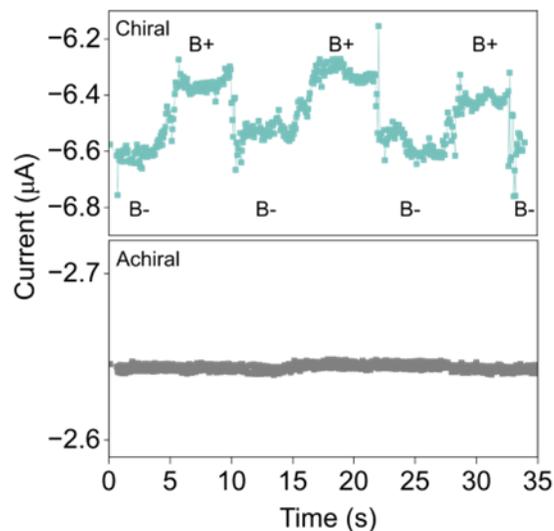

**Supplementary Figure 30.** *I-t* response of chiral (CMHS) and achiral (PEA$_2$PbI$_4$) spin valve at -0.5 V recorded by changing the magnetic field during operation. x-axis scale is adjusted for plotting.